\documentclass[twocolumn,showpacs,preprintnumbers,amsmath,amssymb,aps,pre]{revtex4}

\usepackage{graphicx,color}

\begin{document}

\preprint{}

\title{ Statistical mechanics of Beltrami flows in axisymmetric geometry:\\ Theory reexamined}

\author{Aurore Naso$^1$\footnote{Present address: Laboratoire de Physique, Ecole Normale Sup\'erieure de Lyon and CNRS (UMR 5672), 46, all\'ee d'Italie, 69364 Lyon Cedex 07, France}}
\author{Romain Monchaux$^1$}
\author{Pierre-Henri Chavanis$^2$}
\author{B\'ereng\`ere Dubrulle$^1$}

\affiliation{
$^1$ SPEC/IRAMIS/CEA Saclay, and CNRS (URA 2464), 91191
Gif-sur-Yvette Cedex, France\\
$^2$ Laboratoire de Physique Th\'eorique (UMR 5152), Universit\'e
Paul Sabatier, 118 route de Narbonne,\\ 31062 Toulouse, France\\
}

\begin{abstract}
A simplified thermodynamic approach of the incompressible axisymmetric
Euler equations is considered based on the conservation of helicity,
angular momentum and microscopic energy. Statistical equilibrium
states are obtained by maximizing the Boltzmann entropy under these
sole constraints. We assume that these constraints are selected by the
properties of forcing and dissipation. The fluctuations are found to
be Gaussian while the mean flow is in a Beltrami state. Furthermore,
we show that the maximization of entropy at fixed
helicity, angular momentum and microscopic energy is equivalent to the
minimization of macroscopic energy at fixed helicity and angular
momentum. This provides a justification of this selective decay
principle from statistical mechanics. These theoretical predictions
are in good agreement with experiments of a von K\'arm\'an turbulent
flow and provide a way to measure the temperature of turbulence and
check Fluctuation-Dissipation Relations (FDR). Relaxation equations
are derived that could provide an effective description of the
dynamics towards the Beltrami state and the progressive emergence of a
Gaussian distribution. They can also provide a numerical algorithm to
determine maximum entropy states or minimum energy states.
\end{abstract}

\date{\today}

\maketitle

\section{Introduction}

In a turbulent flow, the number of degrees of freedom scales like
${Re}^{9/4}$, where $Re$ is the Reynolds number, and can reach
$10^{22}$ for atmospheric-like flows, comparable to the Avogadro
number. This is beyond the present capacity of computers. For example,
numerical simulations of a von K\'arm\'an (VK) turbulent flow at
$Re=10^6$, a standard laboratory flow used for turbulence studies (see
below), would require resolutions of the order of $10^{14}$ grid
points and integration times of the order $ 10^5$ years of cpu with
current computers. This conclusion justifies the introduction of
turbulence models to reduce the number of degrees of freedom and make
turbulence amenable to numerical simulation or theoretical
understanding. This goal cannot be reached unless the different
components of turbulence and their interactions are identified.

Turbulence being intrinsically a stochastic process, it can be
decomposed in two components: the mean flow and the fluctuations around
it. A good turbulence model should therefore be able to predict both
the structure of the mean flow, and its influence on and through
fluctuations, within a reduced number of degrees of freedom. This
kind of information is typically provided by statistical mechanics. Can
we adapt statistical methods to deal with the turbulence problem?

This program has been pioneered by Onsager \cite{Onsager}, 
Montgomery \& Joyce \cite{mj} and Lundgren
\& Pointin \cite{lp} in the framework of two-dimensional 
point vortices.  In the last decade, this statistical approach has
been extended by Miller \cite{Miller} and Robert and Sommeria
\cite{Robert_Sommeria_1} to simplified 2D or quasi 2D flows with
continuous vorticity. Even more recently, Leprovost {\it et al.} 
\cite{leprovost} have shown that the 2D formalism could actually be
applied to a typical 2D 1/2 situation, an axisymmetric flow. They
obtained a relationship that gives the general shape of stationary
solutions (mean flows) of the axisymmetric Euler equations. This
relationship has been tested and confirmed experimentally in a
turbulent von K\'arm\'an flow by Monchaux {\it et al.} 
\cite{monchaux1} who observed that, at high Reynolds numbers, the
selected shape is Beltrami, with vorticity and velocity aligned
everywhere. As discussed in Appendix
\ref{sec_lep}, such a shape cannot be obtained with the
thermodynamical approach of Leprovost {\it et al.}
\cite{leprovost}. In the present work, we revisit the theoretical
tools in order to capture Beltrami states as statistical equilibrium
states. We also extend the computations one step further by
considering fluctuations around mean field.

Specifically, we develop a simplified thermodynamic approach based on
the conservation of helicity, angular momentum and microscopic
energy. We assume that these constraints are selected by the
properties of forcing and dissipation. From a maximum entropy
principle we derive the mean flow and the fluctuations around it. We
find that the mean flow is in a Beltrami state and that the
fluctuations are Gaussian.  We also show that {\it the maximization of
entropy at fixed helicity, angular momentum and microscopic energy is
equivalent to the minimization of macroscopic energy at fixed helicity
and angular momentum.} This justifies from statistical mechanics a
selective decay principle introduced previously from phenomenological
arguments \cite{montgomery}. We use the mean field theory to link the
fluctuations to the response of the mean flow to perturbations
(susceptibility) and to the temperature in a way reminiscent of the
Fluctuation-Dissipation Theorem. This provides a way to measure the
temperature of turbulence through the fluctuation level. The analogy
with 2D turbulence is discussed. In fact, due to the dual nature of
axisymmetric flows, intermediate between 2D and 3D turbulence, we find
the emergence of two different effective temperatures in the
fluctuations. One temperature, characterizing velocity fluctuations,
is related to the formation of coherent structures like in 2D
turbulence. Another one, characterizing vorticity fluctuations, is
related to 3D vorticity stretching and diverges with increasing
resolution. These predictions have been tested in companion papers
\cite{monchaux2,papier3} based on PIV measurements in a turbulent von
K\'arm\'an flow and are in fair agreement with observations.

The paper is organized as follows. In  Section \ref{Framework}, we recall the basic problematics associated with statistical mechanics of turbulence, and formulate our hypotheses and the associated theoretical framework.
In Section \ref{Eulersec}, we recall the stationary
solutions and the conservation laws (the backbone
of the statistical mechanics approach) of axisymmetric flows. In Sec. \ref{bel}, we recall the phenomenological selective decay principle leading to Beltrami flows. Section \ref{Stat_mec_gen} is
devoted to the computation of the statistical equilibrium states of axisymmetric inviscid flows using mean
field theory.  We derive the Gibbs states and the
fluctuation-dissipation relations (FDR)  with two different
mean field approximations. In each case, the mean flow is in a Beltrami state and the fluctuations are Gaussian. In Section \ref{SectionConnection}, we make the connection between different variational principles that characterize the equilibrium states. For each principle, we propose a set of relaxation equations that can be used  as a numerical algorithm to solve the variational problem. These relaxation equations can also provide an effective description of the relaxation of the system towards maximum entropy states. Finally, we justify through statistical mechanics the phenomenological principle according to which: ``the mean flow should minimize the macroscopic energy at fixed helicity and angular momentum''.

\section{Hypotheses and theoretical framework}
\label{Framework}

\subsection{Turbulence, Navier-Stokes equations and classical statistical mechanics}

A turbulent flow is described by the Navier-Stokes equations
\begin{equation}
\frac{\partial {\bf u}}{\partial t} +{\bf u}\cdot \nabla {\bf u} =
   -\frac{1}{\rho}\nabla p + \nu \nabla^2 {\bf u}+f,
\label{NS}
\end{equation}
where ${\bf u}$ is the velocity,  $p$ the
pressure, $\rho$ the fluid density, $\nu$ its kinematic viscosity, and $f$ a forcing. In the absence of forcing, the velocity decays to zero due to the dissipation, so that turbulence is an intrinsic out-of-equilibrium problem. In the sequel, we focus on the simplest situation, where forcing and dissipation equilibrate on average, so that stationary states can arise. The goal of the present paper is to describe these stationary states and the fluctuations around them using tools borrowed from classical statistical mechanics. Specifically, we are going to introduce a Hamiltonian system, perform equilibrium or near equilibrium statistical mechanics and compute its equilibrium states.

\subsection{Stationary Navier-Stokes solutions  vs solutions of Euler equations}

Since forcing and dissipation equilibrate on average for stationary solutions of the Navier-Stokes equations, it seems natural to consider this limiting case first in our quest of a  framework suitable for classical statistical mechanics. In such a limit, we get the Euler equations
\begin{equation}
\frac{\partial {\bf u}}{\partial t} +{\bf u}\cdot \nabla {\bf u} =
   -\frac{1}{\rho}\nabla p.
\label{Euler}
\end{equation}
This is indeed a Hamiltonian system as long as one considers regular solutions such as those  based on finite Galerkin expansions. In 2D turbulence, the consideration of Euler solutions to describe Navier-Stokes stationary solutions is well accepted, based on the remark that the vorticity cannot blow up and that the limit $\nu\to 0$ is usually well behaved under reasonable regularity hypothesis. In 3D turbulence, this hypothesis is still controversial since Onsager \cite{Onsager}. One major problem is that one cannot exclude vorticity blows up in 3D, that would make the limit $\nu\to 0$ singular. A signature of this effect is the famous $4/5$ law of homogeneous turbulence that links energy dissipation to the third moment of the velocity increments, independent of any viscosity. For this reason, Onsager \cite{Onsager} suggested to consider weak solutions of the Euler equations to describe stationary states of the Navier-Stokes equations, thereby allowing a finite amount of energy dissipation even in the absence of viscosity. This suggestion was developed recently in an elegant way by Duchon and Robert \cite{RobertDuchon2000}.
However, weak solutions are not directly amenable to methods of classical statistical physics, and nobody has yet succeeded to follow to the end Onsager's suggestion.
In the present case, we overcome this difficulty by considering only regular solutions of the Euler equation. The bonus is that we deal with a Hamiltonian system to which we can apply statistical mechanics. The malus is that we may have lost any connection with actual turbulence. However, in companion papers \cite{monchaux1,monchaux2,papier3} we compare our theoretical predictions with actual experimental turbulent flows and show that they basically agree. In other words, it seems that {\it stationary states and fluctuations of an out-of equilibrium system, the forced Navier-Stokes equations, can be described by statistical equilibrium states and fluctuations of the Euler equations without forcing and dissipation}.

\subsection{The Euler system and conservation laws}

The Euler equations for regular solutions are characterized by a number of conservation laws that depend on the geometry and on the dimension of the system. In 2D turbulence, for example, the conservation laws are the kinetic energy $E=\frac{1}{2}\int {\bf u}^2 d{\bf x}$, the enstrophy $\Omega=\int \omega^2 d{\bf x}$, where $\omega {\bf z}=\nabla\times {\bf u}$ is the vorticity, and, more generally, any function of the vorticity (Casimirs). In 3D turbulence, the generic conserved quantities are the kinetic energy $E$ and the helicity $H=\int {\bf u} \cdot {\boldsymbol \omega}\, d{\bf x}$. Additional conservation laws are possible in the presence of additional symmetries, such as axisymmetry, see \cite{leprovost}.

In the presence of forcing and dissipation, these conservation laws are altered. In the sequel, we shall postulate  that the balance between forcing and dissipation selects some particularly relevant  conservation laws among the infinity of inviscid invariants.  In particular, we shall argue that
there exists relevant situations in which the only conserved quantities are the microscopic energy $E$ and the helicity $H$. This property holds, for example, for very simple solutions of the Euler equations such that the velocity and the vorticity are aligned everywhere in the flow (Beltrami state). Our aim in this paper is not to determine the mechanisms that select these invariants. This is a complicated problem that depends on the properties of forcing and dissipation and on the Reynolds number. However, to motivate our approach, we show in companion papers  \cite{monchaux1,monchaux2,papier3} that our assumptions are consistent with experimental results in the limit of large Reynolds numbers.

\subsection{Boundary conditions}

When we try to model a turbulent flow by numerical simulations, the choice of boundary conditions is a major and complicated problem. Indeed, the results turn out to depend sensitively upon what boundary conditions have been chosen. For example, for two-dimensional Navier-Stokes turbulence, qualitatively
different results have been obtained for free-slip boundary conditions in rigid squares \cite{r1,r2,r3} and rectangles \cite{mj}, for rectangular periodic boundaries \cite{montgomery,r6,r7}, for no-slip circular boundaries \cite{r8,r9}, for free-slip circular boundaries \cite{r10}, for stress-free circular boundaries \cite{r9}, and for no-slip square boundaries \cite{r11}. For our problem, we argue that numerical simulations cannot deal with sufficiently high Reynolds numbers (see the Introduction) and we rather focus on an experimental device, namely a von K\'arm\'an flow \cite{monchaux1,monchaux2}. In that case, the boundary conditions are automatically determined by the experimental geometry and the forcing.

Since our theoretical approach is inviscid, we cannot hope
to describe the experimental flow close to its boundaries. We rather
claim that, when forcing and dissipation equilibrate each other, this
flow can be reasonably described with an inviscid approach far enough
from the boundaries. Our theory will be derived with the inviscid
boundary conditions reflecting a cylindrical closed domain
({\it i.e.} $\psi=0$ on the boundary). The comparison with experiments,
detailed in \cite{papier3}, will be performed over a sub-domain of the
experiment, far from boundary and forcing. As already shown in
\cite{monchaux1,monchaux2}, this restriction is sufficient to obtain
good agreement between inviscid theory and experimental fields, at the
zeroth order approximation (\textit{i.e.} if one considers the mean
flow topology and its small fluctuations).

\section{Euler equation in the axisymmetric case} \label{Eulersec}

\subsection{A convenient formulation of the axisymmetric Euler equations} \label{eq-axi}

In the axisymmetric case, the incompressible Euler equations take the form
\begin{gather}
\frac{1}{r} \frac{\partial}{\partial r} (r u_r) + \frac{\partial u_z}{\partial z}  = 0, \\
\frac{\partial u_r }{\partial t} + u_r \frac{\partial u_r}{\partial r}  + u_z \frac{\partial u_r}{\partial z} -
\frac{u_\theta^2}{r} = -\frac{1}{\rho} \frac{\partial p}{\partial r}, \\
\frac{\partial u_\theta}{\partial t}  + u_r \frac{\partial u_\theta}{\partial r} + u_z \frac{\partial u_\theta}{\partial z}
 + \frac{u_r u_\theta}{r} = 0, \\
\frac{\partial u_z}{\partial t}  + u_r \frac{\partial u_z}{\partial r}  + u_z \frac{\partial u_z }{\partial z} =
-\frac{1}{\rho} \frac{\partial p}{\partial z} ,
\end{gather}
where $(u_r,u_\theta,u_z)$ are the velocity components in a cylindrical referential $(r,\theta,z)$. Here, $r$ runs from $0$ to $R$ and $z$ from $0$ to $2h$ (we take the origin of the $z$ axis at the bottom of the domain). Furthermore, we choose the length unit such that the total volume is unity: $\int rdrdz=1$ (due to the axial symmetry, we systematically divide all the volume integrals by $2\pi$). It was shown in
\cite{leprovost} that the axisymmetric incompressible Euler
equations can be rewritten in a simplified form in terms of $\sigma$,
$\xi$ and $\psi$, where $\sigma=ru_\theta$ is the angular momentum,
$\xi$ is the potential vorticity related to the azimuthal component of the vorticity by
$\xi=\omega_\theta/r$, and $\psi$ is the streamfunction associated with
the poloidal component of the velocity:
\begin{equation}
{\bf u} = u_\theta {\bf
\hat{e}_\theta} + \nabla \times \left( \frac{\psi}{r} {\bf
\hat{e}_\theta} \right).
\label{defivit}
\end{equation}
 Note that $u_r=-\partial_z \psi /r$ and
$u_z=\partial_r \psi /r$. The axisymmetric Euler equations can then
be recast as \cite{leprovost}:
\begin{gather}
\frac{\partial \sigma}{\partial t} + \{ \psi,\sigma \} = 0, \label{sigma} \\
\frac{\partial  \xi}{\partial t} + \{ \psi,\xi \} = \frac{\partial}{\partial z}\left(
\frac{\sigma^2}{4y^2} \right), \label{xi} \\
\Delta_* \psi \equiv \frac{1}{2y} \frac{\partial^2\psi}{\partial{z}^2}+\frac{\partial^2\psi}{\partial{y}^2}= -\xi, \label{psi}
\end{gather}
where $y=r^2/2$, $ \{ \;, \; \}$ is the Poisson bracket ($ \{
\psi,\phi \} =\partial_y \psi \partial_z \phi - \partial_z \psi
\partial_y \phi$) and $\Delta_*$ is a pseudo-Laplacian.

A few general remarks are in order regarding this special case: i) One
sees from Eq. (\ref{sigma}) that the angular momentum is conserved by
the fluid particles and can only be mixed through the Euler
dynamics. This is the analog of vorticity mixing in 2D turbulence and
it justifies the introduction of a mixing entropy for the distribution
of angular momentum (see Sec. \ref{MFPOL}).  We will see that we can
establish a close parallel with the statistical mechanics of 2D
turbulence to determine the distribution of angular momentum. ii) By
contrast, the potential vorticity is stirred like in 3D turbulence and
not conserved. Therefore, the distribution of potential vorticity is
more difficult to investigate and this can lead to complicated
problems such as cascade towards small scales, formation of
singularities etc.  In fact, we shall show in the
companion paper \cite{naso2} that the statistical theory predicts the
existence of large-scale coherent structures (like in 2D) but that
these states are unstable saddle points and should cascade towards
smaller and smaller scales (like in 3D). However, these large-scale
structures can have a very long lifetime because, being saddle points
of entropy, they are unstable only for some particular
perturbations. If the dynamics does not spontaneously generate these
optimal perturbations, the system can remain ``frozen'' in a saddle
point of entropy for a long time \cite{minens,naso2}. Therefore, we
are truly in a situation intermediate between 2D turbulence and 3D
turbulence.

\subsection{Stationary states} \label{statio-axi}

The axisymmetric Euler equations admit an infinite number of steady states. The general form of stationary solutions of the axisymmetric Euler equations (\ref{sigma}-\ref{psi}) has been established in \cite{leprovost}. They are given by
\begin{gather}
\sigma = f(\psi), \label{sigma_stat} \\
- \Delta_* \psi = \xi = \frac{f(\psi)f'(\psi)}{2y}+g(\psi), \label{xi_stat}
\end{gather}
where $f$ and $g$ are arbitrary functions. When $f$ is linear $f(x)=\lambda x$ and $g=0$, the vorticity and velocity are aligned everywhere ${\boldsymbol\omega}= \lambda {\bf u}$ and the stationary flow is a Beltrami state. An alternative form also established in \cite{leprovost} is
\begin{gather}
\psi= R(\sigma), \label{sigma_statbis} \\
 \xi R'(\sigma)- \frac{\sigma}{2y}=Q(\sigma), \label{xi_statbis}
\end{gather}
where $R$ and $Q$ are arbitrary functions. The relation between Eqs. (\ref{sigma_stat}-\ref{xi_stat}) and Eqs. (\ref{sigma_statbis}-\ref{xi_statbis}) is developed in \cite{leprovost} provided that some invertibility properties for the functions are assumed.

\subsection{Conservation laws} \label{cons_axi}

Axisymmetric inviscid flows admit an infinite number of conserved quantities,
namely the total energy
\begin{eqnarray}
E &=&\frac{1}{2} \int (u_r^2+u_\theta^2+u_z^2)rdrdz\nonumber\\
&=&\frac{1}{2} \int \xi
\psi dydz + \frac{1}{4} \int \frac{\sigma^2}{y} dydz, \label{energy-gen}
\end{eqnarray}
the Casimirs
\begin{equation}
I_G = \int G(\sigma) dydz, \label{casimirs}
\end{equation}
and the generalized helicities
\begin{equation}
H_F = \int \xi F(\sigma) dydz, \label{hel_gen}
\end{equation}
where $G$ and $F$ are any (regular) functions.  In the sequel, we also introduce the notation $H_n$ and $I_n$ for the case where $F$ or $G$ are power-laws $x^n$. In particular,  $I=I_1=\int \sigma dydz$ (angular momentum), $\Gamma=H_0=\int \xi dydz$ (circulation) and  $H=H_1=\int \xi \sigma dydz$ (helicity) are conserved.

\subsection{Energy-helicity-Casimir functional} \label{casimir_axi}

From the integral constraints discussed previously, a generalization of the Arnol'd energy-Casimir functional has also been introduced in \cite{leprovost}. This is the energy-helicity-Casimir functional ${A}=E+I_{G}+H_{F}$. Consider the optimization problem
\begin{eqnarray}
\label{i0}
\min_{\xi,\sigma}/\max_{\xi,\sigma}\lbrace \, A[\xi,\sigma] \,\rbrace.
\end{eqnarray}
A critical point of this functional determines a steady state of the axisymmetric equations. Indeed, writing
\begin{equation}
\delta {A}=\delta (E+I_{G}+H_{F})=0,
\label{i1}
\end{equation}
and taking variations on $\sigma$ and $\xi$, we obtain
\begin{equation}
\psi+F(\sigma)=0,
\label{i2}
\end{equation}
\begin{equation}
{\sigma\over 2y}+G'(\sigma)+\xi F'(\sigma)=0,
\label{i3}
\end{equation}
 and we recover the equations (\ref{sigma_statbis}-\ref{xi_statbis}) characterizing a steady solution of the axisymmetric Euler equations. The fact that we obtain all the steady states means that the quantities  given by Eqs. (\ref{energy-gen},\ref{casimirs},\ref{hel_gen}) are the {\it unique} invariants of the axisymmetric incompressible Euler equations \cite{leprovost}. Furthermore, if the critical point of Eq. (\ref{i0})  is a maximum or a minimum of $A$ then this steady state is nonlinearly dynamically stable. In many cases, we shall restrict ourselves to formal nonlinear stability \cite{holm}. We consider {\it small} perturbations and we only require that the critical point  is a (local) maximum or minimum of $A$ such that the second order variations
\begin{eqnarray}
\delta^2 A=\frac{1}{2}\int\delta\xi\delta\psi\, dydz+\int \frac{(\delta\sigma)^2}{4y}\, dydz\nonumber\\
+\frac{1}{2}\int G''(\sigma)(\delta\sigma)^2\, dydz+\frac{1}{2}\int F''(\sigma)\xi (\delta\sigma)^2\, dydz\nonumber\\
+\int F'(\sigma)\delta\xi\delta\sigma\, dydz,
\label{i4}
\end{eqnarray}
are definite positive or definite negative for all perturbations $\delta\sigma$ and $\delta\xi$. Formal stability implies linear stability (in that case $\delta^2A$ can be used as a norm) but it does not imply nonlinear stability for infinite dimensional systems \cite{holm}.

On the other hand, the optimization problem given by Eq. (\ref{i0})  provides just a {\it sufficient} condition of nonlinear dynamical stability. More refined stability conditions can be obtained by adding some constraints in the optimization problem \cite{eht,proc}. For example, the minimization problem
\begin{eqnarray}
\label{gr1}
\min_{\xi,\sigma}\lbrace \, E[\xi,\sigma]\, | \, H_F[\xi,\sigma]=H_F, \, I_G[\xi,\sigma]=I_G \, \rbrace,
\end{eqnarray}
is more refined than
\begin{eqnarray}
\label{gr2}
\min_{\xi,\sigma}\lbrace \, E[\xi,\sigma]+\mu H_F[\xi,\sigma]+\alpha I_G[\xi,\sigma] \, \rbrace,
\end{eqnarray}
in the sense that a solution of Eq. (\ref{gr2}) is always a  solution of the more constrained problem  given by Eq. (\ref{gr1}), but the reciprocal may be  wrong. This is similar to ensemble inequivalence in statistical mechanics  where different ensembles have the same critical points but not necessarily the same maxima or minima (giving rise to different stability criteria) \cite{ellis,proc}. Ensemble inequivalence is generic for systems with long-range interactions like turbulence.

\subsection{Relaxation equations towards dynamical equilibrium}
\label{algo}

We can introduce a set of relaxation equations that solve the optimization problem given by Eq. (\ref{i0}) by  adapting the general methods described in \cite{pre,proc}. We write the relaxation equations as
\begin{eqnarray}
{\partial\xi\over\partial t}=X,  \qquad
{\partial \sigma\over\partial t}=Y.
\label{algo1}
\end{eqnarray}
The time variations of ${A}$ are given by
\begin{eqnarray}
&&\dot A=\int  X \left(\psi+F(\sigma)\right)dydz\nonumber\\
&&+\int  Y \left(\frac{\sigma}{2y}+G'(\sigma)+\xi F'(\sigma)\right)dydz.
\label{algo2}
\end{eqnarray}
To determine the functions $X$ and $Y$, we maximize the rate of production (resp. dissipation) of ${A}$ with the constraints
\begin{eqnarray}
{X^{2}\over 2}\le C_{\xi}, \qquad
{Y^{2}\over 2}\le C_{\sigma}.\label{algo3}
\end{eqnarray}
This is the counterpart of Onsager's linear thermodynamics. The variational principle can be written in the form
\begin{eqnarray}
\delta \dot{A}
+\int \frac{1}{\chi} \delta \left(\frac{X^2}{2} \right) dydz
+\int \frac{1}{D} \delta \left( \frac{Y^2}{2} \right) dydz = 0,\nonumber\\
\label{algo4}
\end{eqnarray}
where $\chi$ and $D$ are Lagrange multipliers associated with the constraints given by Eqs. (\ref{algo3}).
This leads to the relaxation equations
\begin{eqnarray}
\frac{\partial \xi}{\partial t}=X=-\chi\left\lbrack\psi+F(\sigma)\right\rbrack=-\chi \frac{\delta A}{\delta\xi},\label{algo5}
\end{eqnarray}
\begin{eqnarray}
\frac{\partial\sigma}{\partial t}=Y=-D\biggl\lbrack {\sigma\over 2y}+G'(\sigma)+\xi F'(\sigma)\biggr \rbrack=-D \frac{\delta A}{\delta\sigma}.\label{algo6}
\end{eqnarray}
It is straightforward to establish that
\begin{eqnarray}
\dot A=-\int\frac{X^2}{\chi}\, dydz-\int\frac{Y^2}{D}\, dydz.\label{algo7}
\end{eqnarray}
Therefore, the relaxation equations (\ref{algo5},\ref{algo6}) satisfy $\dot {A}\le 0$ if $D$ and $\chi$ are both positive and $\dot {A}\ge 0$ if $D$ and $\chi$ are both negative. On the other hand, $\dot {A}= 0$ iff $X=Y=0$ so that $(\xi,\sigma)$ is a steady state. By Lyapunov's direct method, we conclude that these equations can only converge
towards a maximum of $A$ (if $D$, $\chi$ are negative) or a minimum of ${A}$ (if $D$, $\chi$ are positive). Saddle points of $A$ are linearly unstable. Therefore, the relaxation equations (\ref{algo5},\ref{algo6})  can be used as a numerical algorithm to solve the optimization  problem given by Eq. (\ref{i0}).

\section{Beltrami flows}\label{bel}

Let us consider the minimization of energy at fixed helicity and angular momentum \footnote{In this paper, we shall not take into account the conservation of circulation $\Gamma=\int \xi\, dydz$ at equilibrium because there is no critical point of energy at fixed helicity, angular momentum and circulation: if the Lagrange multiplier $\gamma$ associated with the conservation of $\Gamma$ is non-zero, the differential equation resulting from the variational principle $\delta E-\mu\delta H-\alpha\delta I-\gamma\delta\Gamma=0$ presents some divergencies at $r=0$ (see \cite{naso2}). However, in Appendix \ref{sec_other}, we present dynamical equations that dissipate energy at fixed helicity, angular momentum and circulation.}
\begin{eqnarray}
\label{bel1}
\min_{\xi,\sigma} \lbrace \, E[\xi,\sigma]\, | \, H, \, I \, \rbrace.
\end{eqnarray}
The critical points of this variational principle satisfy
\begin{eqnarray}
\label{bel2}
\delta E+\mu \delta H +\alpha \delta I=0,
\end{eqnarray}
where $\mu$ and $\alpha$ are Lagrange multipliers.
Taking the variations of $\xi$ and $\sigma$, we obtain
\begin{gather}
\psi+\mu\sigma=0, \label{bel3} \\
\frac{\sigma}{2y}+\mu\xi+\alpha=0.
\end{gather}
These equations  can be rearranged in the form
\begin{gather}
\sigma = -\frac{1}{\mu} \psi, \label{bel4} \\
- \Delta_* \psi = \xi = \frac{\psi}{2\mu^2 y}-\frac{\alpha}{\mu}. \label{bel5}
\end{gather}
They define a steady state of the axisymmetric Euler equations of the form (\ref{sigma_stat},\ref{xi_stat}) with  $f(x)=-x/\mu$ linear and $g(x)=-\alpha/\mu$ constant. In that case, the vorticity and the (relative) velocity are aligned everywhere
\begin{eqnarray}
\label{bel6}
{\boldsymbol\omega}= -\frac{1}{\mu} ({\bf u}+\alpha{\bf e}_z\times {\bf r}),
\end{eqnarray}
and the stationary flow is a Beltrami state. This critical point is a (local) minimum of energy at fixed helicity and angular momentum iff
\begin{eqnarray}
\frac{1}{2}\int\delta\xi\delta\psi\, dydz+\int \frac{(\delta\sigma)^2}{4y}\, dydz
+\mu\int \delta\xi\delta\sigma\, dydz\ge 0, \nonumber\\
\label{bel7}
\end{eqnarray}
for all perturbations $\delta\sigma$ and $\delta\xi$ that conserve helicity and angular momentum at first order (see Appendix \ref{A1}).

The variational problem given by Eq. (\ref{bel1}) can be given several justifications:

(i) It can be introduced in a
phenomenological manner from a {\it selective decay principle}
\cite{montgomery}. Due to a small viscosity, or other dissipative
or relaxation mechanisms, the energy (fragile invariant) is dissipated
while helicity and angular momentum (robust invariants) are
approximately conserved. This selective decay principle has a long
history in physics. It first appeared in the MHD literature with
Taylor's explanation of some behavior of the Zeta reversed-field pinch
due to a conjectured rapid decay of magnetic energy relative to
magnetic helicity \cite{taylor}. This principle leads to a force-free
state, {\it i.e.} a state whose magnetic field is proportional to its own
curl
\cite{woltjer,mtv}. This is an analog  of the Beltrami states that
hydrodynamicists subsequently discovered in connection with
axisymmetric turbulence. They are also related to minimum enstrophy
states in 2D turbulence introduced by Bretherton \& Haidvogel
\cite{bh} and later by Leith
\cite{leith}, leading to linear relationship between vorticity and
stream function. Using the Chandrasekhar-Kendall eigenfunctions of the
curl, these Beltrami states are easy to construct in both MHD and
hydrodynamics. For example, they were used as a Galerkin basis for an
extensive set of turbulent MHD computations by Shan {\it et al.} 
\cite{s1,s2}.

(ii) In Sec. \ref{SectionConnection}, we shall propose a
justification of the minimization problem given by Eq. (\ref{bel1})
based on statistical mechanics arguments. To our knowledge, this
statistical mechanics justification has not been given before.

(iii) According to
Eq. (\ref{gr1}), the minimization problem given by Eq. (\ref{bel1})
-if it has a solution- determines a steady state of the axisymmetric
Euler equations that is formally nonlinearly stable.

{\it Remark:} the minimization problem given by Eq. (\ref{bel1}) may not have a solution, {\it i.e.}  a minimum of energy at fixed helicity and angular momentum may not exist. This is the conclusion that we shall reach in \cite{naso2}. The absence of equilibrium state is usually associated with a ``collapse" like the gravothermal catastrophe or the isothermal collapse in self-gravitating systems \cite{paddy,ijmpb}. In the present context, the ``collapse'' is associated with the break-up of large scale structures and the cascade of energy at smaller and smaller scales. The relaxation equations associated with the minimization problem given by Eq. (\ref{bel1}), derived in Sec. \ref{sec_relaxbel},  may give a qualitative idea of how the system evolves by dissipating energy \cite{naso2}. However, since these equations are purely phenomenological, we stress that they may not necessarily provide  an accurate description of the true evolution of the system.

\section{Statistical mechanics of the axisymmetric Euler-Beltrami system}\label{Stat_mec_gen}

\subsection{Basic set-up}

In the previous sections, we considered steady states of the axisymmetric Euler equations. In a realistic situation where the system is forced and dissipated at small scales, these steady states describe the mean flow resulting from the balance between forcing and dissipation. However, there also exists fluctuations around the mean flow so that the
velocity field is ${\bf u}=\overline{\bf u}+{\bf u}'$, where $\overline{\bf u}$ is the averaged velocity field and ${\bf u}'$ the fluctuations. We shall assume that $\overline{\bf u}$ is axisymmetric and that the total system evolves while conserving the energy $E=\int \overline{{\bf u}^2}\, d{\bf r}$, the helicity $H=\int \overline{{\bf u}\cdot {\boldsymbol \omega}}\, d{\bf r}$ and the angular momentum $I=\int r\overline{u_\theta}\, d{\bf r}$, but no other constraint. We assume that these conservation laws are selected by the properties of forcing and dissipation, and consequently by the Reynolds numbers. We shall call such flows an Euler-Beltrami system. It was found experimentally \cite{monchaux1,monchaux2}  that the system approaches a Beltrami state when the Reynolds number is sufficiently large, giving support to the basic assumption of our theory.

In the sequel, it will prove useful to operate a poloidal/toroidal decomposition such that ${\bf u}_p=(u_r, 0,u_z)$ and ${\bf u}_t=(0,u_\theta,0)$. In term of these fields, the kinetic energy density ${\bf u}^2={\bf u}_p^2+{\bf u}_t^2$ while the helicity density ${\bf u}\cdot{\boldsymbol \omega}={\bf u}_p\cdot{\boldsymbol \omega}_p+{\bf u}_t\cdot {\boldsymbol \omega}_t$. For axisymmetric fields, $\int {\bf u}_p\cdot{\boldsymbol \omega}_p\, d{\bf r}=\int {\bf u}_t\cdot  {\boldsymbol \omega}_t\, d{\bf r}$. To determine the distribution of angular momentum $\sigma$  and vorticity $\xi$,  we shall use a Mean Field
Theory (MFT). This method is traditionally very efficient in systems
of high dimensionality, or with long-range interactions, a condition met in
fluid mechanics. In our system, we have at our disposal two privileged directions: the toroidal  direction and the poloidal direction. We therefore derive two different MFT procedures, freezing the fluctuations in one of the two directions to capture the fluctuations in the other direction. In each case, we introduce a suitable entropy, and maximize it under the energy, helicity and angular momentum constraints so as to obtain the Gibbs states. From these Gibbs states, we derive relations for the mean flow and for the fluctuations.  The first approach, which is closely related to the approach in 2D turbulence will give us the mean field (Beltrami) and the distribution of angular momentum (Gaussian). It will allow us to justify a principle of minimum energy at fixed helicity and angular momentum. The second approach will give us the same mean field and the distribution of vorticity (Gaussian). Fluctuations are however found to diverge in the limit of number of modes going to infinity, a pathology that can be traced back to vorticity stretching.

\subsection{Mean field approximation on the poloidal field: the distribution of angular momentum $\sigma$}
\label{MFPOL}

\subsubsection{Computations}

Let us first assume that the fluctuations are mainly in the toroidal direction, so that the poloidal fluctuations can be ignored  $|{\bf u}'_p|\ll |\overline{\bf u}_p|$. In that case, the poloidal field is only determined by $\xi=\overline{\xi}$ and $\psi$, so that ${\bf u}'$ is made only by fluctuations of $\sigma$. Let us introduce the density probability $\rho({\bf
r},\eta)$ to measure $\sigma=\eta$ at position ${\bf
r}=(y,z)$. Then, the local moments of the angular momentum are $\overline{\sigma^n}=\int \rho\eta^n\, d\eta$. To proceed further, we need to introduce an entropy. Since the angular momentum density $\sigma$ is conserved by the flow but undergoes a complicated mixing process (like the vorticity in 2D), it is natural to introduce the mixing entropy
\begin{equation}
S[\rho] = - \int \rho \ln \rho \, dydzd\eta,
\label{mixing}
\end{equation}
similar to the one introduced by Miller-Robert-Sommeria in 2D turbulence. We expect the entropy to increase during  the dynamics (while the helicity, the angular momentum  and the microscopic energy are conserved) until the flow achieves a steady state. The functional given by Eq. (\ref{mixing})  can also be interpreted as the neg-information (the opposite of the information). Maximizing this neg-information under given constraints is the simplest procedure we can adopt to compute the fluctuations, according to the information theory and its application to statistical mechanics developed by Jaynes \cite{Jaynes57}.

In our approach, the conserved quantities are
\begin{eqnarray}
{E}^{f.g.}&=&\frac{1}{2}\int {\overline{\xi} \; {\psi}} \, dydz+\int \frac{\overline{\sigma^2}}{4y} \, dydz\nonumber\\
&=&\frac{1}{2}\int {\overline{\xi} \; {\psi}} \, dydz+\int \rho \frac{\eta^2}{4y} \, dydzd\eta,
\label{energy_xi} \\
{H} &=& \int \overline{\xi} \, \overline{\sigma}\, dydz= \int \overline{\xi}  \rho\eta\,  dydzd\eta,
\label{hel_gen_xi}\\
{I} &=& \int \overline{\sigma}\, dydz= \int   \rho\eta\,  dydzd\eta.
\label{add1}
\end{eqnarray}
The first constraint given by Eq. (\ref{energy_xi}) will be called  the microscopic (or fine-grained) energy because it takes into account the fluctuations of $\sigma$. It is different from the macroscopic (or coarse-grained) energy
\begin{eqnarray}
{E}^{c.g.}=\frac{1}{2}\int {\overline{\xi} \; {\psi}} \, dydz+\int \frac{\overline{\sigma}^2}{4y}  \, dydz,
\label{ecg}
\end{eqnarray}
which ignores these fluctuations. We have ${E}^{f.g.}={E}^{c.g.}+E_{fluct}$. In our terminology, the energy will be called a {\it fragile constraint} because it cannot be expressed in terms of the coarse-grained field since $\overline{\sigma^2}\neq \overline{\sigma}^2$. While the microscopic energy ${E}^{f.g.}$ is conserved, the macroscopic energy ${E}^{c.g.}$ is not conserved and can decay. There is the same distinction between the fine-grained enstrophy ${\Gamma}_2^{f.g.}=\int \overline{\omega^2}\, d{\bf r}$ and the coarse-grained enstrophy ${\Gamma}_2^{c.g.}=\int \overline{\omega}^2\, d{\bf r}$ in 2D turbulence \cite{jfm1,minens}. On the other hand, the helicity given by Eq. (\ref{hel_gen_xi}) and the angular momentum given by Eq. (\ref{add1}) will be called  {\it robust constraints} because they can be expressed in terms of the coarse-grained fields. We shall come back to this important distinction in Sec. \ref{SectionConnection}.

The most probable distribution at metaequilibrium is obtained by
maximizing the mixing entropy $S[\rho]$ at fixed ${E}^{f.g.}$,
${H}$, $I$  and local normalization $\int \rho
d\eta = 1$. Introducing Lagrange multipliers, the
variational principle can be written as
\begin{eqnarray}
\delta S - \beta_\xi \delta {E}^{f.g.}  - \mu_\xi \delta {H}-\alpha_\xi\delta I\nonumber\\
-\int \zeta({\bf r})\delta\left (\int \rho d\eta\right )\, d{\bf r}= 0,
\end{eqnarray}
where $\beta_\xi$ is the inverse temperature and $\mu_\xi$ the helical potential (we have written these quantities with a subscript  $\xi$ to recall that the fluctuations of $\xi$ are ignored in the present approach).
The variations on $\overline{\xi}$ imply
\begin{equation}
\beta_\xi {\psi} + \mu_\xi \overline{\sigma} = 0, \label{xi_moy}
\end{equation}
while the variations on $\rho$ yield the Gibbs state
\begin{equation}
\rho({\bf r},\eta) = \frac{1}{Z} e^{-\frac{\beta_\xi
\eta^2}{4y}-(\mu_\xi \overline{\xi}+\alpha_\xi)\eta},
\label{xi_fluc}
\end{equation}
where the ``partition function'' is determined via the normalization condition
\begin{equation}
Z({\bf r}) = \int e^{-\frac{\beta_\xi\eta^2}{4y}-(\mu_\xi\overline{\xi}+\alpha_\xi)\eta} d\eta.
\end{equation}
From Eq. (\ref{xi_fluc}), the local average  of the angular momentum is
\begin{gather}
\overline{\sigma}=-\frac{2y}{\beta_\xi}(\mu_\xi \overline{\xi}+\alpha_\xi).
\label{Bel_sigma_avg}
\end{gather}
Together with Eq. (\ref{xi_moy}), this equation determines a Beltrami state. On the other hand $\rho({\bf r},\eta)$, the distribution of the
fluctuations of $\sigma$, is Gaussian with centered variance
\begin{gather}
\sigma_2 \equiv
\overline{\sigma^2}-\overline{\sigma}^2=\frac{2y}{\beta_\xi}.
\label{Bel_sigma_var}
\end{gather}
The Gibbs state can be rewritten
\begin{equation}
\rho({\bf r},\eta) = \left (\frac{\beta_\xi}{4\pi y}\right )^{1/2} e^{-\frac{\beta_\xi}{4y}(\eta-\overline{\sigma})^2}.
\label{xi_flucbis}
\end{equation}
Therefore, our statistical theory based on the conservation of ${E}^{f.g.}$,
${H}$ and $I$  predicts that the mean flow is a Beltrami state with Gaussian fluctuations of angular momentum.

Note that  Eq. (\ref{Bel_sigma_var}) means that the toroidal velocity fluctuations are uniform
\begin{equation}
\overline{u_\theta^2}-\overline{u_\theta}^2=\frac{1}{\beta_\xi}.
\label{Bel_sigma_var-ut}
\end{equation}
Therefore, $\beta_\xi$ can be interpreted  as an inverse temperature measuring the fluctuations of ${u}_\theta$. These predictions
enable the measurements of effective temperatures of turbulence
through fluctuations of $u_{\theta}$ in a
Beltrami flow. Because variances are positive,
$\beta_\xi$ is always positive, unlike in the 2D
situation where the temperature can be negative (in the present context, the inverse temperature is the equivalent of the Lagrange multiplier associated with the conservation of microscopic enstrophy in 2D turbulence, which is positive \cite{minens}) . Note that Eq.~(\ref{Bel_sigma_var-ut})
predicts {\it uniformity} of azimuthal velocity fluctuations which is an interesting prediction of our theory. This has been confirmed experimentally in \cite{monchaux2}.

On the other hand, the energy contained in the fluctuations is simply
\begin{eqnarray}
E_{fluct}=\int\rho \, \frac{(\eta-\overline{\sigma})^2}{4y} \,  dydzd\eta=\int \frac{\sigma_2}{4y}\, dydz=\frac{1}{2\beta_\xi}.\nonumber\\ \label{energy_fluc2}
\end{eqnarray}
 Therefore, the statistical temperature $\beta_\xi^{-1}$ can also be interpreted as the energy (by unit volume) of the toroidal fluctuations. Moreover,  for simple  Beltrami flows with $\alpha_\xi=0$,   there is equipartition between the macroscopic energy in the poloidal and toroidal directions
\begin{equation}
{E_p}^{c.g.}\equiv \int \frac{\overline{\xi} \; {\psi}}{2} \, dydz=
\int \frac{\overline{\sigma}^2}{4y} dydz\equiv {E_t}^{c.g.},
\label{equipart1}
\end{equation}
with a simple connection with the helicity as
\begin{equation}
{E_p}^{c.g.}=-\frac{\mu_\xi}{\beta_\xi}{H}.
\label{relat}
\end{equation}

\subsubsection{Comments}
\label{Comments}

The statistical equilibrium state given by Eqs.~(\ref{xi_moy}) and (\ref{Bel_sigma_avg}) is of the form of Eqs.~(\ref{sigma_stat}) and (\ref{xi_stat}) where $f$  is linear $f(x)=\lambda x$ (with $\lambda=-\beta_\xi/\mu_{\xi}$) and $g$ is constant (with $g=-\alpha_\xi/\mu_{\xi}$). This means that the equilibrium state is a stationary solution  of the axisymmetric Euler equation and takes the shape of a Beltrami state (see Eq. (\ref{bel6})).

We can also provide an interesting interpretation of our fluctuation relation Eq.~(\ref{Bel_sigma_var-ut}),
predicting uniformity of azimuthal velocity fluctuations. This equation shows that the azimuthal
velocity fluctuations define an effective statistical temperature
$1/\beta_{\xi}$. This equation may be
regarded as formally analogous to a Fluctuation Dissipation Relation (FDR) since it links
fluctuations and temperature. These predictions
enable the measurements of turbulence effective temperatures
through fluctuations of $u_{\theta}$ in a
Beltrami flow. As discussed previously,
$\beta_\xi$ is always positive. In contrast,
$\mu_\xi$ can take positive or negative values,
depending on the helicity sign.

The analogy between our predictions and FDRs can actually be
pushed forward. Indeed, another possible way to derive Eq.~(\ref{Bel_sigma_var-ut}) is to introduce, as in classical statistical
mechanics, the partition function $Z$ describing the
Beltrami equilibrium state in the mean field approximation:
\begin{equation}
\overline{\sigma^2}-\overline{\sigma}^2 =
\frac{1}{\mu_{\xi}^2}\frac{\delta^2 \log Z}{\delta
\overline{\xi}^2}= -\frac{1}{\mu_{\xi}}\frac{\delta
\overline{\sigma}}{\delta \overline{\xi}},\label{FDRterun}
\end{equation}
where $\delta$ stands for functional derivative. Formally, the mathematical object $\delta
\overline{\sigma}/\delta \overline{\xi}$ can be seen as a response
function. With this point of view, Eq.~(\ref{FDRterun}) again
reflects a formal analogy with FDRs since another classical way to
write it down is to link the fluctuations of a field to its
response to a perturbation.

\subsection{Mean field approximation in the toroidal direction: the distribution of vorticity $\xi$}

\subsubsection{Spectral approach}

We now assume that the fluctuations in the toroidal direction are frozen so that $\sigma=\overline{\sigma}$, and that they do not depend on the azimuthal direction.
Since the vorticity $\xi$ is {\it not} conserved, we cannot in principle rigorously apply a statistical mechanics to the fluctuations of vorticity. We present here a phenomenological approach, based on neg-information rather than mixing entropy, and will test its relevance by comparison with experimental data in companion papers \cite{monchaux2,papier3}. 

In our approach, the conserved quantities are
\begin{eqnarray}
E&=&\frac{1}{2}\int \overline{\psi\xi}  dydz + \int
\frac{\overline{\sigma}^2}{4y} dydz, \label{energy_sig} \\
H &=& \int \overline{\sigma}\, \overline{\xi} dydz.
\label{hel_gen_sig}\\
{I} &=& \int \overline{\sigma}\, dydz.
\label{add2}
\end{eqnarray}
In the expression of the energy, we note that the fluctuations of angular momentum have been neglected so that $\overline{\sigma^2}=\overline{\sigma}^2$ while the fluctuations of potential vorticity have been taken into account so that $\overline{\psi\xi}\neq \overline{\psi}\, \overline{\xi}$.  In order to deal with the term $\overline{\psi\xi}$ that introduces a nonlocality, we shall develop the statistical theory in the  spectral space by using an approach similar to that developed by Kraichnan \cite{kraichnan} and Salmon {\it et al.} \cite{salmon} in 2D turbulence.  Let us first note that the field $\phi=\psi/r$ satisfies the differential equation
\begin{eqnarray}
{\cal L}\phi\equiv -\Delta\phi+\frac{1}{r^2}\phi=r\xi=\omega_\theta.
\label{calL}
\end{eqnarray}
To solve the problem, we decompose the fields onto the eigenfunctions ${\phi_{mn}}$ of ${\cal L}$ defined by
\begin{equation}
{\cal L} {\phi}_{mn}\equiv -\Delta {\phi}_{mn}+\frac{\phi_{mn}}{r^2}=B_{mn}^2 \phi_{mn},
\label{eigenmodes}
\end{equation}
with $\phi_{mn}=0$ on the boundary. Taking the origin of the $z$ axis at the bottom of the domain, the eigenfunctions are given by the Hankel-Fourier modes
\begin{equation}
{\phi}_{mn}=\sqrt{\frac{2}{h R^2 J_2(j_{1m})^2}}J_1\left(\frac{j_{1m}r}{R}\right)\sin\left( \frac{n\pi z}{2h}\right),
\end{equation}
where $j_{1m}$ is the $m^{th}$ zero of Bessel function $J_1$. The mode $\phi_{mn}$ corresponds to $m$ cells in the radial direction and $n$ cells in the vertical direction. The corresponding eigenvalues are
\begin{equation}
B_{mn}^2=  \left(\frac{j_{1m}}{R}\right)^2+ \left( \frac{n\pi}{2h} \right)^2.
\end{equation}
The eigenfunctions are orthogonal with respect to the scalar product
\begin{eqnarray}
\left\langle f g \right\rangle\equiv \int_0^R\int_{0}^{2h}r\,dr dz\, f g,
\label{prod_sc}
\end{eqnarray}
so that $\left\langle {\phi}_{mn} \phi_{m'n'} \right\rangle=\delta_{mm'}\delta_{nn'}$ where $\delta_{ij}$ is the Kronecker symbol. We now decompose the fields on these eigenmodes writing
\begin{gather}
\frac{\psi}{r}=\phi=\sum_{m=1}^{N_m}\sum_{n=0}^{N_n} \Psi_{mn}{\phi}_{mn}, \label{psi_def} \\
\omega_\theta=r\xi=\sum_{m=1}^{N_m}\sum_{n=0}^{N_n} \omega_{mn}{\phi}_{mn}, \label{xi_def} \\
{u}_\theta=\frac{\sigma}{r}=\sum_{m=1}^{N_m}\sum_{n=0}^{N_n} u_{mn}{\phi}_{mn}, \label{sig_def}
\end{gather}
where we have restricted the sum over finite number of modes, so as to respect Hamiltonian condition for the Euler equation. Moreover, the finite number of modes implies a coarse graining of the solution, that is desirable to reach a stationary state. With this decomposition, using Eq. (\ref{calL}), we have by construction
\begin{equation}
\omega_{mn}=B_{mn}^2\Psi_{mn}. \label{rel_Bel}
\end{equation}
Inserting the decomposition given by Eqs. (\ref{psi_def}-\ref{sig_def}) in the expression of the  energy and helicity, and using the orthogonality condition given by Eq. (\ref{prod_sc}) and the Parseval identities, we obtain
\begin{gather}
E=\frac{1}{2} \sum_{mn} \left( \frac{\overline{\omega_{mn}^2}}{B_{mn}^2}+u_{mn}^2 \right),\\
H= \sum_{mn}\omega_{mn}u_{mn},\\
I= \sum_{mn}u_{mn}\langle r \phi_{mn}\rangle,
\end{gather}
where the brackets denote a domain average. To apply the statistical theory, we introduce the density probability $\rho_{mn}(\nu)$ of measuring the value $\omega_{mn}=\nu$ of the vorticity in the mode $(m,n)$. We can rewrite the constraints in the form
\begin{gather}
E=\frac{1}{2}\sum_{mn}\int \left( \frac{\nu^2}{B_{mn}^2}+u_{mn}^2 \right) \rho_{mn}(\nu)\, d\nu,\\
H=\sum_{mn}\int \omega_{mn}u_{mn}\rho_{mn}(\nu)\, d\nu,\\
I= \sum_{mn}u_{mn}\langle r \phi_{mn}\rangle.
\end{gather}

\subsubsection{The multi-modes case}

We first consider the situation where the energy is spread all over the different wavenumbers $(n,p)$. In that case, there is no reason to specialize a special mode and we can use the neg-information in spectral space to define the entropy
\begin{equation}
S=-  \sum_{mn}\int\rho_{mn}(\nu)\ln\rho_{mn}(\nu)\, d\nu.
\end{equation}
The Gibbs state is obtained by maximizing $S$ at fixed energy, helicity, angular momentum and normalization $\int \rho_{mn}\, d\nu=1$. We write the variational principle as
\begin{eqnarray}
\delta S-\beta_\sigma \delta E-\mu_\sigma \delta H -\alpha_\sigma\delta I\nonumber\\
- \sum_{mn}\chi_{mn}\delta \left (\int \rho_{mn}d\nu\right ) =0,
\end{eqnarray}
where $\beta_\sigma$, $\mu_\sigma$ and $\alpha_\sigma$  are Lagrange multipliers (the subscript $\sigma$ recalls that the fluctuations of angular momentum have been neglected).  The variations on $u_{mn}$ yield
\begin{equation}
\beta_\sigma u_{mn}+\mu_\sigma\omega_{mn}+\alpha_\sigma\langle r \phi_{mn}\rangle=0.
\end{equation}
Multiplying by $\phi_{mn}$, summing on the modes and using $r=\sum_{mn}\langle r \phi_{mn}\rangle \phi_{mn}$, we obtain
\begin{equation}
\frac{\beta_\sigma \overline{\sigma}}{2y}+ \mu_\sigma
\overline{\xi}+\alpha_\sigma=0.
\label{sigma_moy1}
\end{equation}
The variations on $\rho_{mn}$ yield the Gibbs state
\begin{equation}
\rho_{mn}(\nu)=\frac{1}{Z_{mn}} e^{-\left(\frac{\beta_\sigma\nu^2}{2B_{mn}^2}+\mu_\sigma u_{mn}\nu \right) },
\label{gibbstor}
\end{equation}
where $Z_{mn}$ is a factor ensuring the local normalization condition. The distribution of the fluctuations of $\omega_{mn}$ is therefore Gaussian with
\begin{gather}
\beta_\sigma\frac{\overline{\omega_{mn}}}{B_{mn}^2}+\mu_\sigma u_{mn}=0,\label{bof1}\\
\overline{\omega_{mn}^2}-\overline{\omega_{mn}}^2=\frac{B_{mn}^2}{\beta_\sigma}.
\label{bof}
\end{gather}
According to Eq. (\ref{bof}), the mean fluctuating energy per mode number $(\overline{\omega_{mn}^2}-\overline{\omega_{mn}})^2/B_{mn}^2$ is constant: we have equipartition of energy for the poloidal fluctuations. This result is a classical outcome of equilibrium statistical mechanics.
Multiplying Eq. (\ref{bof1}) by $\phi_{mn}$, summing on the modes and using Eq. (\ref{rel_Bel}), we obtain
\begin{equation}
\beta_\sigma\overline{\psi}+\mu_\sigma \overline{\sigma}=0.
\label{stationsigma}
\end{equation}
Equations (\ref{sigma_moy1}) and (\ref{stationsigma}) show that the mean flow associated with the statistical equilibrium state is a stationary solution of the axisymmetric Euler equations corresponding to a Beltrami state. One can also deduce from Eq. (\ref{bof}) that  the total volumic energy contained in the fluctuations is
\begin{equation}
E_{fluct}=\sum_{mn}
\int \frac{(\nu-\overline{\omega_{mn}})^2}{2B_{mn}^2} \rho_{mn}(\nu) \, d\nu=\frac{N_{tot}}{2\beta_\sigma}, \label{energy_fluctor}
\end{equation}
where $N_{tot}$ is the total number of modes. The linear divergence
with $N_{tot}$ comes from energy equipartition.  Therefore, the
statistical temperature in this case is proportional to the poloidal
energy of fluctuations. This is therefore analog to the previous mean
field case.

\begin{figure}[t]
\includegraphics[width=4.2cm]{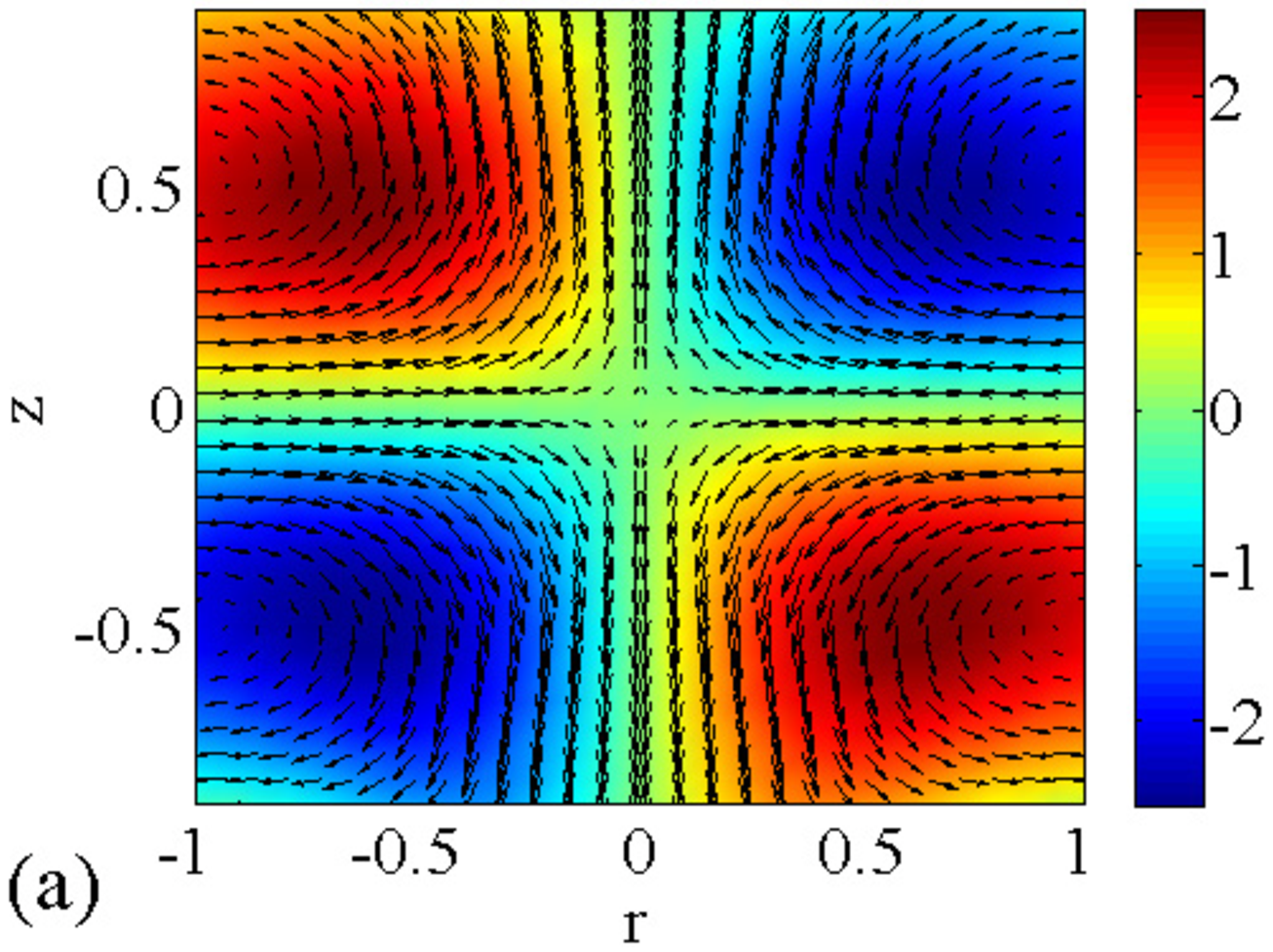}
\includegraphics[width=4.2cm]{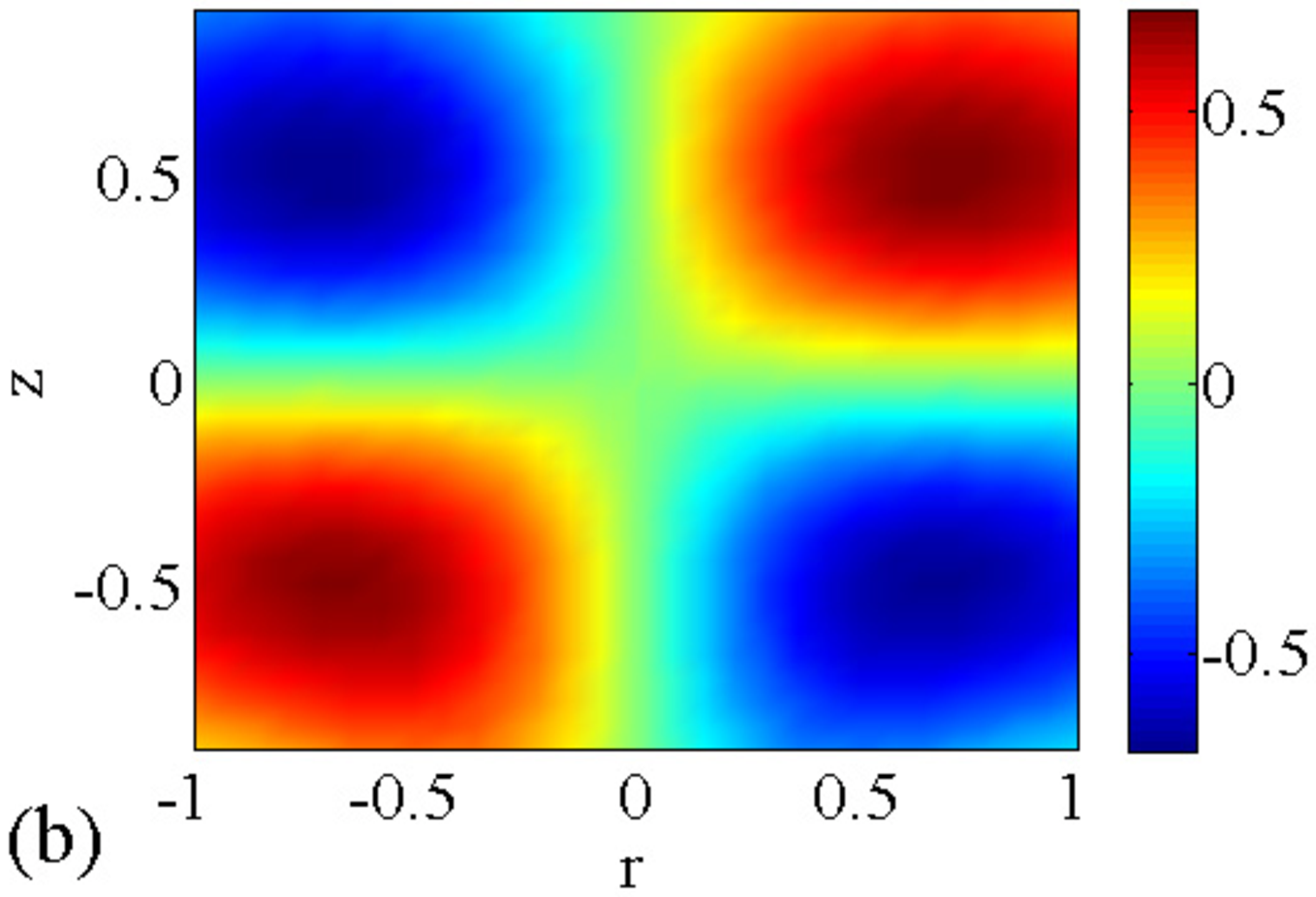}
\includegraphics[width=4.2cm]{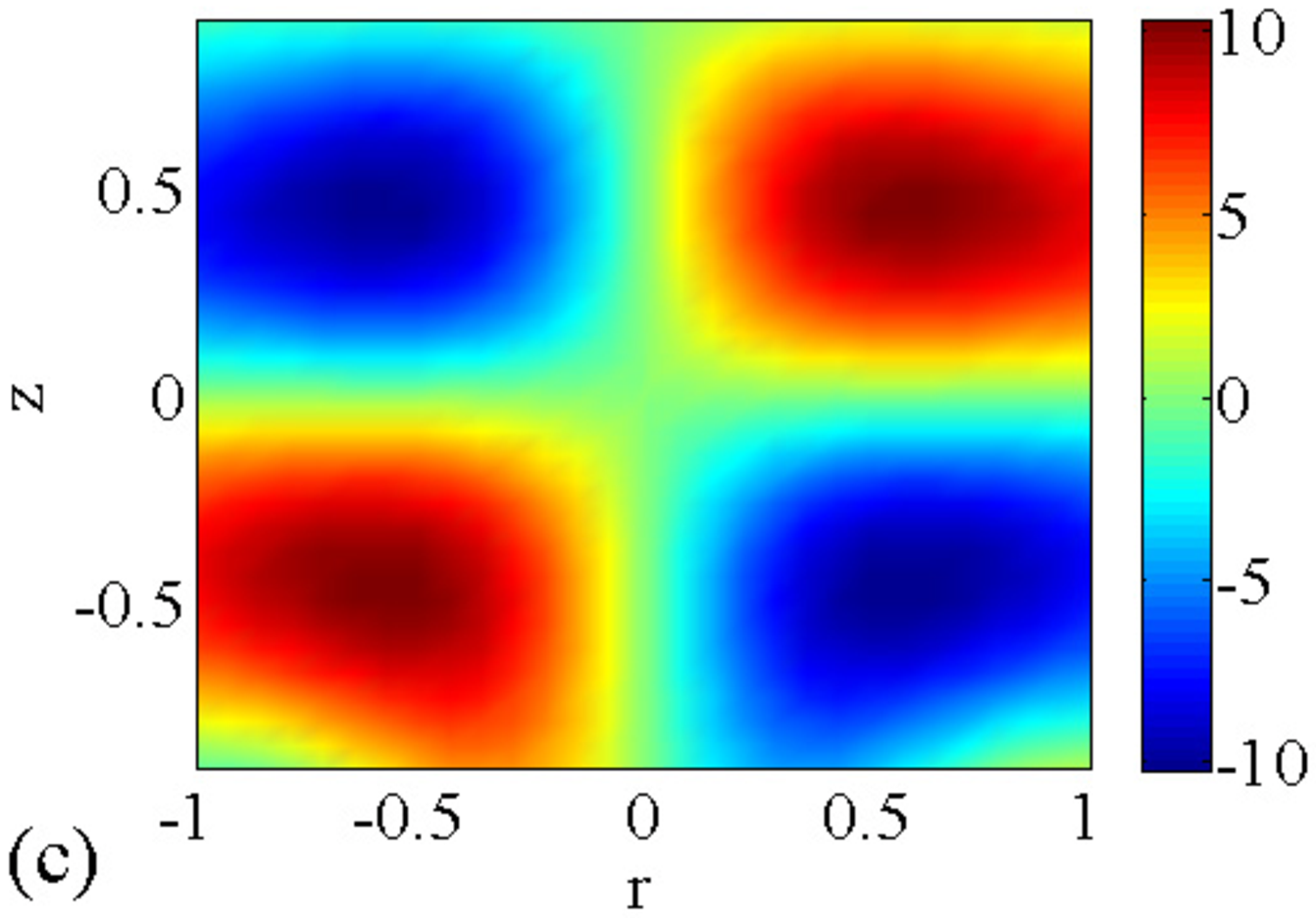}
\includegraphics[width=4.2cm]{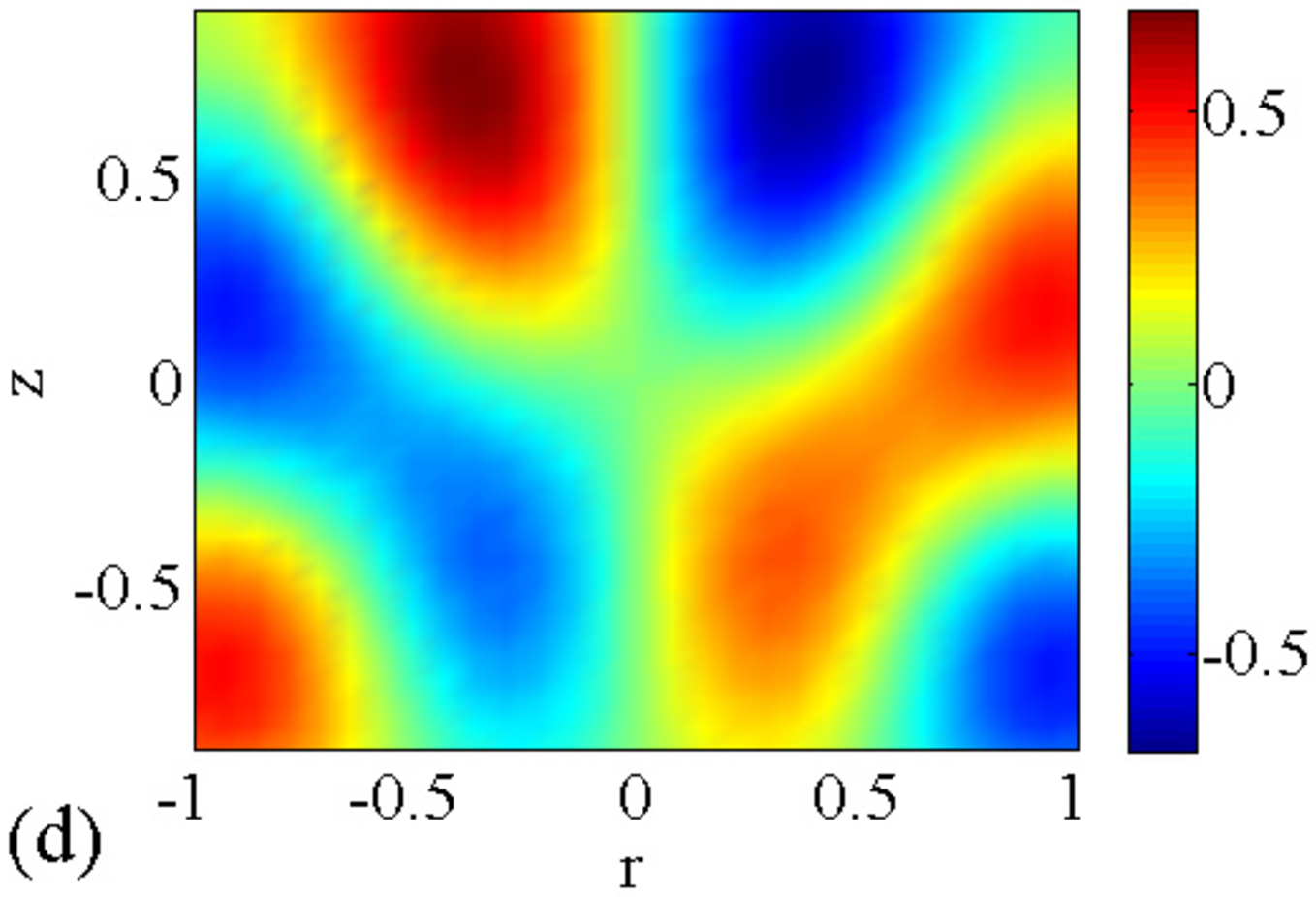}
\caption{(Color online) Example of synthetic instantaneous fields in the multi-mode case, at $-\beta_\sigma/\mu_\sigma=-3.6$ and $\beta_\sigma=2$: (a) projection of $(\overline{u_r},\overline{u_\theta},\overline{u_z})$ over a vertical meridional plane. The toroidal velocity is coded in color, the poloidal velocity is denoted with arrows; (b) projection of $\overline{\phi}$ over a vertical meridional plane; (c) projection of $\overline{\omega_\theta}$ over a vertical meridional plane; (d) projection of $\omega'_\theta$ over a vertical meridional plane. The field is constructed with $N_m=2$ radial modes and $N_m=2$ axial modes. The amplitude of the velocity modes results from a least square fit to the average velocity field described in \cite{monchaux2}, with a forcing using TM73 turbines in contra-rotation at $F=6Hz$ over eigenmodes given by Eqs. (\ref{eigenmodes}). The instantaneous value is obtained by drawing one realization of random vorticity fluctuations according to the theoretical Gibbs distribution given by Eq. (\ref{gibbstor}).}
\label{fig:dess}
\end{figure}

\begin{figure}[t]
\includegraphics[width=4.2cm]{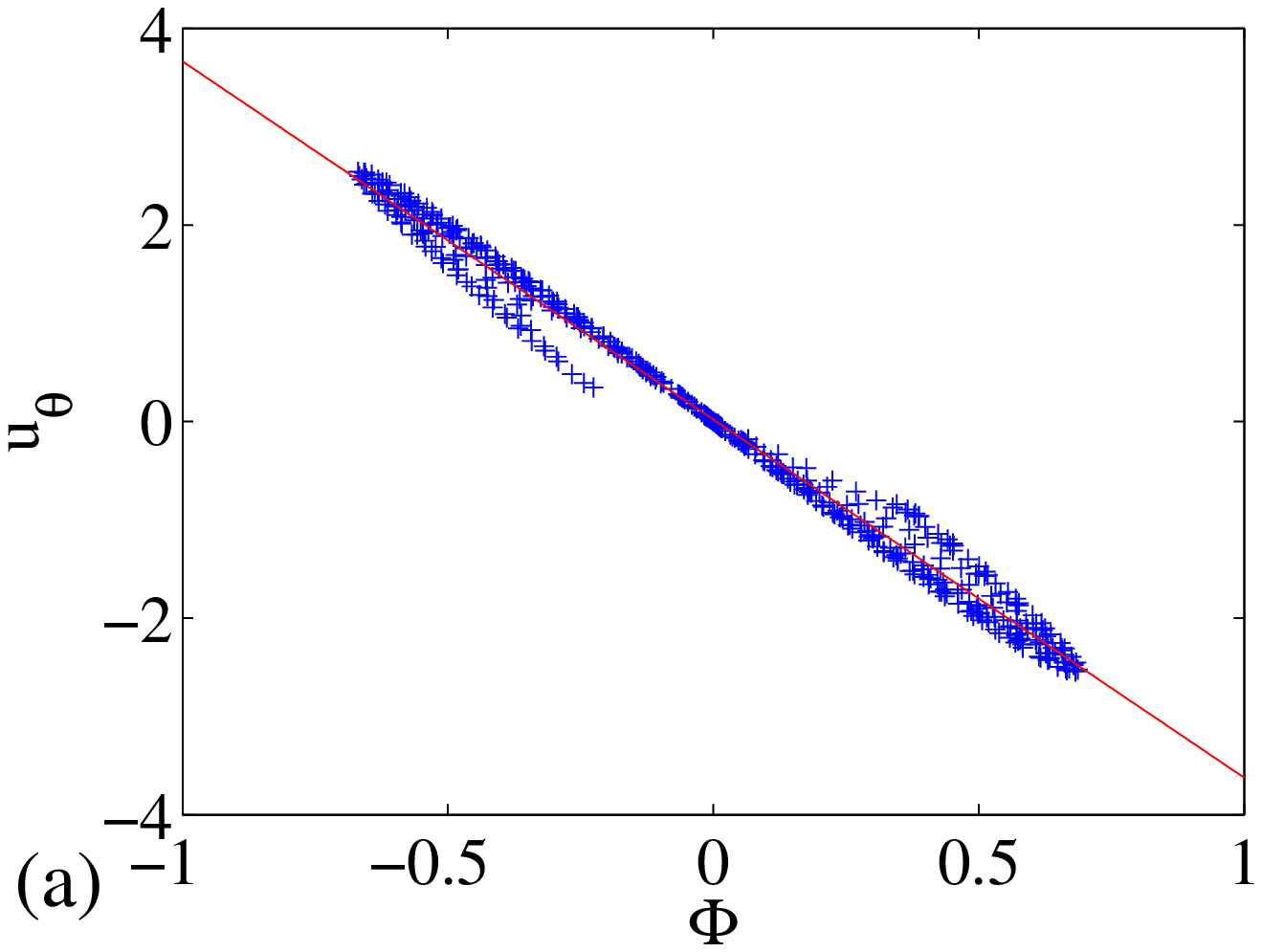}
\includegraphics[width=4.2cm]{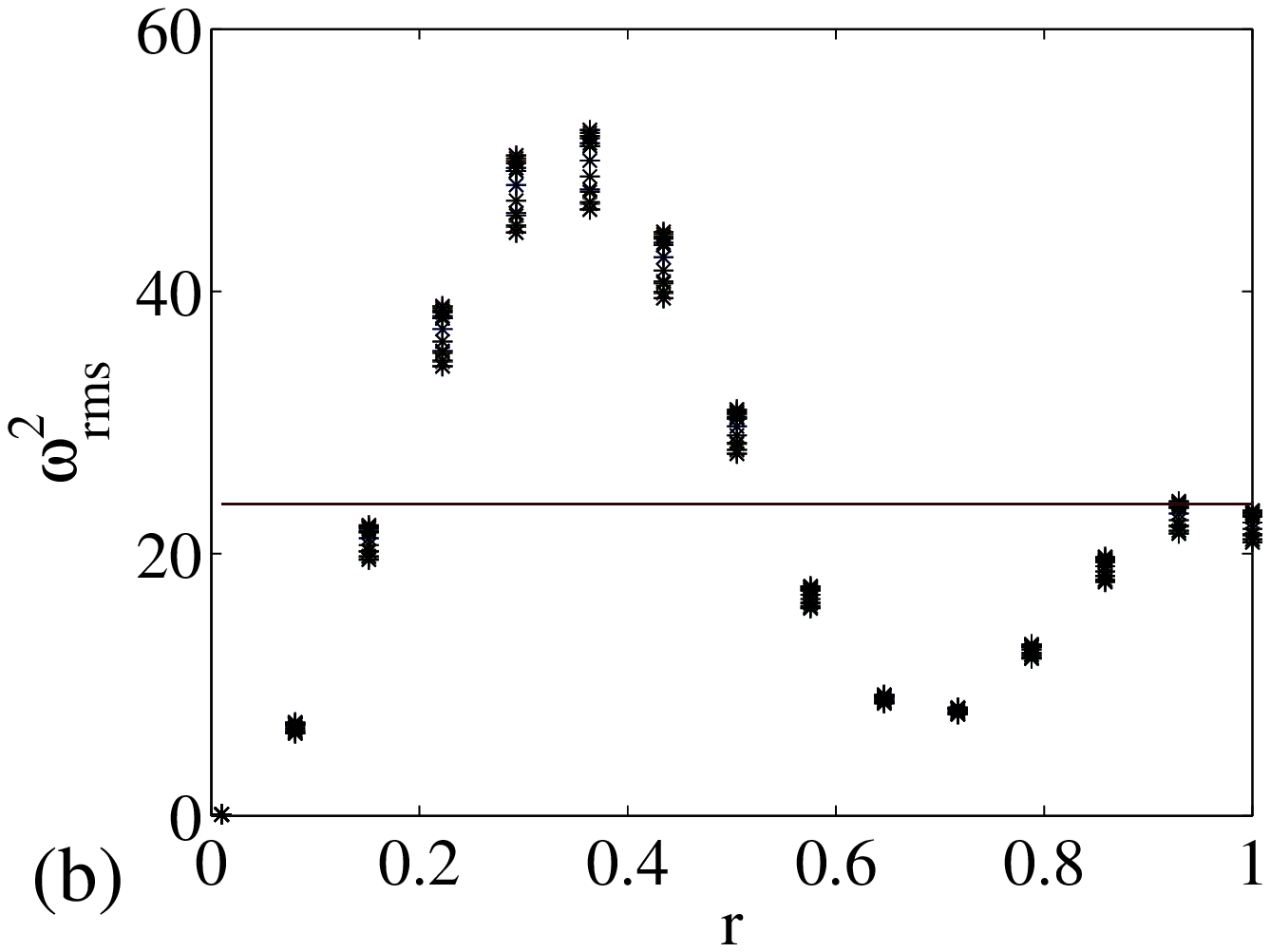}
\caption{(a) (Color online) Toroidal velocity $u_\theta$ as a function of the stream function $\phi$.
The line is a linear fit, with slope $B=-3.6$. (b) Spatial
variation of the theoretical vorticity variance as a function of $r$
only. This variance has been obtained using the modes of the Hankel
decomposition of the velocity field and Eq. (\ref{vorti1}).  The
dispersion in vertical direction corresponds to a dispersion of
variance at a given $r$ with varying $z$. The horizontal line is the
empirical value $N_{tot}\langle B^2\rangle/\beta_\sigma$. The
conditions are the same as in Fig. \ref{fig:dess}.}
\label{fig:dess2}
\end{figure}

It is also interesting to compute the azimuthal vorticity fluctuations. They are given by
\begin{equation}
\overline{\omega_\theta^2}-\overline{\omega_\theta}^2=\frac{1}{\beta_\sigma }\sum_{mn}B_{mn}^2 {\phi}_{mn}^2,
\label{vorti1}
\end{equation}
since $\omega_{mn}$ and $\omega_{m'n'}$
are independent if $(m,n) \neq (m',n')$.
For comparison with real turbulent data fields, we have constructed synthetic instantaneous fields obeying  Eqs. (\ref{sigma_moy1}),  (\ref{gibbstor}) and (\ref{stationsigma}) to study some of their properties. An example is shown in  Fig. \ref{fig:dess}. The average velocity field is obtained as follows: we choose a given number of modes to represent the field $(N_m,N_n)$. Then, we get the coefficient of the Hankel-decomposition $u_{mn}$ by a least-square fit to $\overline{u_\theta}$ of an actual mean turbulent field (here, the velocity field obtained by counter-rotation at $F=6Hz$ of TM73 impellers-this field is described in \cite{monchaux2}). This field is shown in Fig. \ref{fig:dess}. Once the $u_{mn}$ are obtained, we get the coefficient $\Psi_{mn}$ of the Hankel-decomposition of the mean stream function $\phi$ from a least-square fit to the poloidal experimental velocity fields $\overline{u_r}$ and $\overline{u_z}$ using Eqs. (\ref{defivit}). Such a field, shown in Fig. \ref{fig:dess},  obeys the relation 
(\ref{stationsigma}) with $B=-\beta_\sigma/\mu_\sigma=-3.6$, as shown in Fig. \ref{fig:dess2}. We then obtain  the mean azimuthal vorticity $\omega_\theta$ thanks to (\ref{calL}). It is shown in Fig. \ref{fig:dess}. We also use the modes of the decomposition to compute the theoretical variance $\omega_{rms}^2$ following Eq. (\ref{vorti1}), shown in Fig. \ref{fig:dess2}. Finally, we compute an instantaneous fluctuation field of azimuthal vorticity by drawing for each mode a realization of the Gaussian distribution Eq. (\ref{gibbstor}) and reconstructing the field through Eq. (\ref{xi_def}).
The results are provided in Fig. \ref{fig:dess}. We see that the theoretical variance $\omega_{rms}^2$ is not independent of $r$ and $z$ but its dependence in $z$ is weak. In the radial direction, it oscillates mildly around a value $\Omega_\infty$:
\begin{equation}
\overline{\omega_\theta^2}-\overline{\omega_\theta}^2\approx \Omega_\infty.
\label{mild}
\end{equation}
Noteworthy, $\Omega_{\infty}$ is much larger than the value
$B^2/\beta_\sigma$ obtained in the one mode case (see
Eq. (\ref{vorti}) of Sec. \ref{sec_onem}). This is reminiscent of what
has been observed in real turbulent data fields \cite{monchaux2}. It
is therefore interesting to study further the dependence of
$\Omega_\infty$ with respect to the number of modes of the problem. 
We have found empirically  that this value  behaves like
\begin{equation}
\Omega_\infty= N_{tot}\frac{\langle B^2\rangle}{\beta_\sigma},
\label{omegainfty}
\end{equation}
where $\langle B^2\rangle$ is a mean Beltrami factor, defined as:
\begin{equation}
\langle B^2\rangle\equiv \frac{1}{N_{tot}}\sum_{mn}B_{mn}^2 .
\label{meanbeltra}
\end{equation}
This number depends on the set of modes (values of $n$ and $m$) considered. For example, in an isotropic situation  when one sums only over modes such that $n=m$, $B_{mn}^2\sim n^2$ and $\langle B^2\rangle\sim N_{tot}^{-1}N_n^3\sim N_{tot}^{1/2}$ (here $N_{tot}=N_nN_m=N_n^2$). In an anisotropic situation such as $N_m=1$, $B_{mn}^2\sim n^2$ and $\langle B^2\rangle\sim N_{tot}^{-1}N_n^3\sim N_{tot}^{2}$ (here $N_{tot}=N_nN_m=N_n$).

Integrating Eq. (\ref{vorti1}) over the volume, we get the azimuthal enstrophy fluctuations
\begin{eqnarray}
\Omega_{fluct}=
\int (\overline{\omega_\theta^2}-\overline{\omega_\theta}^2) dydz=\frac{1}{\beta_\sigma}\sum_{mn}B_{mn}^2= \frac{N_{tot}}{\beta_\sigma}\langle B^2\rangle . \nonumber\\
\label{enstrophy_fluctor}
\end{eqnarray}
Since $\langle B^2\rangle$ grows algebraically with $N_{tot}$ (see
above), the enstrophy fluctuations therefore diverge as $N_{tot}\times
N_{tot}^{\alpha}$ when the number of modes becomes infinite. The
linear part of the divergence is comparable with the linear divergence
obtained for the variance of energy fluctuations, and can be thought
to be an outcome of equilibrium statistical mechanics. Note that in
real turbulent flows, -out-of-equilibrium solutions with non-zero
energy flux-, the energy of fluctuations by mode $E_{fluct}/N_{tot}$
remains finite but the enstrophy of fluctuations by mode
$\Omega_{fluct}/N_{tot}\sim N_{tot}^{\alpha}$ diverges algebraically
with the number of modes in 3D, while the divergence is much milder
(logarithmic) in 2D. This difference can be seen as the signature of
3D vortex stretching, that is captured by our model. Indeed, vortex
stretching induces a transfer of vorticity at continuously decreasing
scales, thereby leading to an enstrophy divergence.

Note that both the enstrophy fluctuations and the energy fluctuations provide an estimate of the statistical temperature. The comparison of temperature in between the two measurements actually provides a measure of the number of modes in the system since energy fluctuations are proportional to $N_{tot}$ and enstrophy fluctuations behave roughly like $N_{tot}^{1.5}$ for an isotropic situation. This will be further discussed in Sec. \ref{discupolo}.

\subsubsection{The one mode case}
\label{sec_onem}

The previous Gibbs distributions do not couple modes with different wavenumbers. They are the analogs of the Boltzmann laws found in simple quasi-geostrophic models \cite{salmon} in the case where energy is spread evenly over all modes (an implicit assumption behind our choice of entropy). It is interesting to consider the opposite case,  where all the energy is concentrated in one mode $(m_c,n_c)$. The results pertaining to this case have been discussed in \cite{monchaux2}. We present here the corresponding detailed computation. In such a case, it is easy to check that
\begin{equation}
\psi=\frac{2y\xi}{B^2}+\kappa y,
\label{equilibr1}
\end{equation}
where we have noted $B\equiv B_{m_c n_c}$ for brevity. Furthermore, the term $\kappa y$ can always be introduced since it is in the kernel of $\Delta$. Since the energy is concentrated in one mode in the spectral space, we cannot use the neg-information anymore in that space to determine the Gibbs distribution. However, by the Linquist theorem, such a peaked probability in the spectral space corresponds to a spread probability in the physical space. We therefore turn back to the density probability  $\rho({\bf r},\nu)$ to measure $\xi=\nu$ at position ${\bf r}$, introduce the neg-information
\begin{eqnarray}
S[\rho] = - \int \rho \ln \rho \, dydzd\nu,
\end{eqnarray}
in the physical space, and consider the maximization of $S[\rho]$ at fixed ${E}$, ${H}$ and normalization $\int \rho\, d\nu=1$. Using Eq. (\ref{equilibr1}), the constraints  can  be written
\begin{eqnarray}
E&=&\int \frac{y}{B^2}\overline{\xi^2} \, dydz + \frac{\kappa}{2}\int \overline{\xi}y \, dydz+\int
\frac{\overline{\sigma}^2}{4y} \, dydz \nonumber\\
&=& \int \frac{y}{B^2}\rho \nu^2\, dydzd\nu +\frac{\kappa}{2}\int y\rho\nu \, dydzd\nu+\int
\frac{\overline{\sigma}^2}{4y} \, dydz,\nonumber\\
\end{eqnarray}
\begin{eqnarray}
H = \int  \overline{\sigma}\, \overline{\xi}\, dydz=\int  \overline{\sigma}\rho\nu\, dydzd\nu,
\end{eqnarray}\begin{eqnarray}
I =  \int  \overline{\sigma}\, dydz.
\end{eqnarray}
We write the variational principle as
\begin{eqnarray}
\delta S-\beta_\sigma \delta E-\mu_\sigma \delta H-\alpha_\sigma\delta I\nonumber\\
- \int\chi({\bf r})\delta \left (\int \rho\, d\nu\right )\, d{\bf r}=0.
\end{eqnarray}
The variations on $\overline{\sigma}$ imply
\begin{equation}
\frac{\beta_\sigma \overline{\sigma}}{2y} + \mu_\sigma \overline{\xi}+\alpha_\sigma= 0, \label{sigma_moy}
\end{equation}
and the variations on $\rho$ yield the Gibbs state
\begin{equation}
\rho({\bf r},\nu) = \frac{1}{Z} e^{-\frac{\beta_\sigma y}{B^2}\nu^2-(\mu_\sigma \overline{\sigma}+\frac{1}{2}\beta_\sigma \kappa y)\nu}, \label{sigma_fluc}
\end{equation}
where $Z$ is the normalization factor. The distribution is therefore Gaussian. Its first two moments are
\begin{eqnarray}
\overline{\xi}&=&-B^2\frac{\mu_\sigma}{\beta_\sigma}\frac{\overline{\sigma}}{2y}-\frac{B^2\kappa}{4},\label{Bel_xi_avg} \\
\xi_2 &\equiv& \overline{\xi^2}-\overline{\xi}^2=\frac{B^2}{2y \beta_\sigma}. \label{Bel_xi_var}
\end{eqnarray}
The Gibbs state can be rewritten
\begin{equation}
\rho({\bf r},\nu) = \left (\frac{y\beta_\sigma}{\pi B^2}\right )^{1/2} e^{-\frac{y\beta_\sigma}{B^2}(\nu-\overline{\xi})^2}.
\end{equation}
One sees that relations (\ref{Bel_xi_avg}) and (\ref{sigma_moy}) can be satisfied simultaneously only if $B^2=(\beta_\sigma/\mu_\sigma)^2$ and $\kappa=4\alpha_\sigma\mu_\sigma/\beta_\sigma^2$. This fixes the wavenumbers $(m_c,n_c)$  of the mode in which energy is accumulated and provides a physical interpretation of the ratio $\beta_\sigma/\mu_\sigma$. The relations (\ref{Bel_xi_avg}) and (\ref{equilibr1}) show that the mean flow associated with the  statistical equilibrium state is a stationary solution of the axisymmetric Euler equation corresponding to a Beltrami state. Moreover, Eq. (\ref{Bel_xi_var}) shows that the fluctuations of vorticity are uniform and scale as
\begin{equation}
\overline{\omega_\theta^2}-\overline{\omega_\theta}^2=\frac{B^2}{\beta_\sigma }.
\label{vorti}
\end{equation}
Integrating over the volume, we find that the azimuthal enstrophy fluctuations are
\begin{equation}
\Omega_{fluct}=
\int (\overline{\omega_\theta^2}-\overline{\omega_\theta}^2)\, dydz= \frac{1}{\beta_\sigma}B^2. \label{enstrophy_fluctoronemode}
\end{equation}
We may also compute the energy contained in the fluctuations as
\begin{eqnarray}
E_{fluct}=\int \frac{y(\nu-\overline{\xi})^2}{B^2} \rho dydzd\eta=\int \frac{y}{B^2}\xi_2 \, dydz=\frac{1}{2\beta_\sigma}. \label{energy_fluc}\nonumber\\
\end{eqnarray}
 Therefore, the statistical temperature can be simply interpreted here as the volumic energy of the poloidal fluctuations. Note that in such a simple case, $\Omega_{fluct}/E_{fluct}=B^2$, consistent with  the multi-mode result since $B^2=\langle B^2\rangle$ in the ``one mode case". Moreover, for simple  Beltrami flows with $\alpha_\sigma=0$, there is equipartition between the energy of the mean flow in the poloidal $E_p$ and toroidal direction $E_t$:
\begin{equation}
E_p^{c.g.}\equiv \frac{1}{2}\int  \overline{\xi}\, \overline{\psi}\, dydz=
\int \frac{\overline{\sigma}^2}{4y} \, dydz\equiv E_t^{c.g.},
\label{equipart}
\end{equation}
with a simple connection with the mean helicity as
\begin{equation}
E_p^{c.g.}=-\frac{\mu_\sigma}{2 \beta_\sigma}{H}.
\label{relat2}
\end{equation}
This case is therefore the analog to the mean field toroidal case, with uniformity of fluctuations and simple connection with helicity.

\subsubsection{A note on vorticity fluctuations}
\label{discupolo}

The link of the present results with experiments has been partially discussed in \cite{monchaux1,monchaux2}, with focus on the mean field toroidal case and the one-mode mean field poloidal case. A detailed comparison is provided in a companion paper \cite{papier3}. It is however interesting to come back to one puzzling result
of \cite{monchaux2} to give it a new interpretation in the present
context.  It has indeed been found that in a turbulent
counter-rotating von K\'arm\'an flow both the velocity and vorticity
fluctuations are approximately uniform over the box.  When interpreted
in the context of the mean field toroidal case, and the one-mode mean
field poloidal case, the value of this constant provides an estimate
of the two inverse temperatures $\beta_\xi$ and $\beta_\sigma$ through
Eqs. (\ref{Bel_sigma_var-ut}) and (\ref{vorti}). Experimentally, one
finds $\beta_\xi\ll \beta_\sigma$, with a ratio
$\beta_\xi/\beta_\sigma$ ranging from $8$ to $17$ in different forcing
configurations. In the restricted context discussed in
\cite{monchaux2}, this difference is puzzling, and points towards the
existence of two different temperatures. If however one considers a
wider context, in which the vorticity fluctuations are considered to
span several modes, the experimental measurements allow for another
interpretation. Indeed, using our least-square fitting to experimental
data with varying $N_n$ and $N_m$ with $N_n=N_m$ (isotropic case), we
found that
$\overline{\omega_\theta^2}-\overline{\omega_\theta}^2\approx
N_{tot}^{1.5}\frac{3.6^2}{\beta_\sigma}$, instead of the value
$3.6^2/\beta_\sigma$ predicted in the one-mode case $N_{tot}=1$
(recall that $B=-3.6$ in the selected experimental
data).  Assuming $\beta_\xi=\beta_\sigma$ instead of $N_{tot}=1$, we
can then use the experimental measurements to infer $N_{tot}$. One
finds a value ranging from $N_{tot}=4$ to $N_{tot}=6$. Such a small
number is intriguing because the usual belief is that turbulent flows
are characterized by a very large number of degrees of freedom. There
are however other indications ({\it e.g.} in the dynamo context \cite{P4})
that turbulent flows can be described using tools adapted from
dynamical systems, as if the {\sl effective} number of degrees of
freedom were indeed small.

\section{Connection between different variational principles}
\label{SectionConnection}

In this section, we make the connection between different variational
principles that characterize the equilibrium states. For each
principle, we propose a relaxation equation (see Appendix
\ref{sec_rela}) that can be used as a numerical algorithm to solve the
corresponding variational problem. These relaxation equations can also
provide an effective description of the relaxation of the system
towards the equilibrium state.  They will be solved numerically in
\cite{papier3}.   We finally justify through statistical
mechanics the phenomenological principle according to which: ``the
mean flow should minimize the macroscopic energy at fixed helicity and
angular momentum''. In this section, we ignore the fluctuations of
vorticity and exclusively consider the mean field theory of the
poloidal field developed in Sec. \ref{MFPOL}.

\subsection{The basic variational principle}
\label{BasicPrin}

The basic maximization problem that we have to solve is:
\begin{equation}
\max_{\rho,\overline{\xi}}\lbrace S[\rho]\, | \, E^{f.g.}, \, H,\, I, \, \int \rho d\eta=1  \rbrace,
\label{bes1}
\end{equation}
with
\begin{equation}
S[\rho] = - \int \rho \ln \rho \, dydzd\eta,\label{bes2}
\end{equation}
and
\begin{eqnarray}
{E}^{f.g.}&=&\frac{1}{2}\int {\overline{\xi} \; {\psi}} \, dydz+\int \frac{\overline{\sigma^2}}{4y} \, dydz\label{bes3}\nonumber\\
&=&\frac{1}{2}\int {\overline{\xi} \; {\psi}} \, dydz+\int \rho \frac{\eta^2}{4y} \, dydzd\eta,\label{bes4}
 \\
{H} &=& \int \overline{\xi} \, \overline{\sigma}\, dydz= \int \overline{\xi}  \rho\eta\,  dydzd\eta,\label{bes5}
\\
{I} &=& \int  \overline{\sigma}\, dydz= \int \rho\eta\,  dydzd\eta.\label{bes6}
\end{eqnarray}
The critical points are determined by the variational principle
\begin{eqnarray}
\delta S - \beta \delta {E}^{f.g.}  - \mu \delta {H} -\alpha \delta I\nonumber\\
-\int \zeta({\bf r})\delta\left (\int \rho d\eta\right )\, dydz= 0.\label{bes7}
\end{eqnarray}
The variations on $\overline{\xi}$ imply
\begin{equation}
\beta {\psi} + \mu \overline{\sigma} = 0,\label{bes8}
\end{equation}
while the variations on $\rho$ yield the Gibbs state
\begin{equation}
\rho({\bf r},\eta) = \left (\frac{\beta}{4\pi y}\right )^{1/2} e^{-\frac{\beta}{4y}(\eta-\overline{\sigma})^2},\label{bes9}
\end{equation}
with
\begin{gather}
\overline{\sigma}=-\frac{2y}{\beta}( \mu \overline{\xi}+\alpha),\label{bes10}\\
\sigma_2 \equiv
\overline{\sigma^2}-\overline{\sigma}^2=\frac{2y}{\beta}.\label{bes11}
\end{gather}
As already indicated, the last relation shows that the temperature must be positive.
Furthermore, a critical point of (\ref{bes1}) is an entropy {\it maximum} at fixed $E^{f.g.}$, $H$, $I$ and normalization  iff
\begin{eqnarray}
\delta^2 J\equiv -\frac{1}{2}\int \frac{(\delta\rho)^2}{\rho}\, dydzd\eta\nonumber\\
-\frac{\beta}{2}\int\delta\overline{\xi}\delta\psi\, dydz-\mu\int\delta\overline{\xi}\delta\overline{\sigma}\, dydz\le 0,
\label{bes12}
\end{eqnarray}
for all perturbations $\delta\rho$ and $\delta\overline{\xi}$ that conserve energy, helicity, angular momentum and normalization at first order .

\subsection{An equivalent but simpler variational principle}
\label{simplerprin}

The maximization problem (\ref{bes1}) is difficult to solve because the stability condition (\ref{bes12}) is expressed in terms of the  distribution $\rho({\bf r},\eta)$. We shall here introduce an equivalent but simpler maximization problem by ``projecting'' the distribution on a smaller subspace. To solve the maximization problem (\ref{bes1}), we can proceed in two steps \cite{proc}.

(i) {\it First step:} we first maximize $S$ at fixed $E^{f.g.}$, $H$, $I$, $\int \rho\, d\eta=1$ {\it and} $\overline{\sigma}({\bf r})=\int\rho\eta\, d\eta$ and $\overline{\xi}({\bf r})$. Since the specification of $\overline{\sigma}({\bf r})$ and $\overline{\xi}({\bf r})$ determines $\int \psi\overline{\xi}\, d{\bf r}$, $H$ and $I$, this is equivalent to maximizing $S$ at fixed $\int \rho \frac{\eta^2}{4y}\, dydzd\eta$, $\int \rho\, d\eta=1$ {and} $\overline{\sigma}({\bf r})=\int\rho\eta\, d\eta$. Writing the variational problem as
\begin{eqnarray}
\delta S-\beta\delta \left (\int \rho \frac{\eta^2}{4y}\, dydzd\eta\right )-\int\lambda({\bf r})\delta\left (\int\rho\eta\, d\eta\right )\, dydz\nonumber\\
-\int \zeta({\bf r})\delta\left (\int\rho\, d\eta\right )\, dydz=0,\qquad\label{ses1}
\end{eqnarray}
we obtain
\begin{equation}
\rho_1({\bf r},\eta) = \left (\frac{\beta}{4\pi y}\right )^{1/2} e^{-\frac{\beta}{4y}(\eta-\overline{\sigma})^2},\label{ses2}
\end{equation}
and we check that it is a global entropy maximum with the previous constraints since $\delta^2 S=-\int\frac{(\delta\rho)^2}{2\rho}\, dydzd\eta\le 0$ (the constraints are linear in $\rho$ so their second variations vanish). We also note that the centered local variance of this distribution $\rho_1$ is
\begin{gather}
\sigma_2 \equiv
\overline{\sigma^2}-\overline{\sigma}^2=\frac{2y}{\beta},\label{ses3}
\end{gather}
implying $\beta\ge 0$.
Using the optimal distribution given by Eq. (\ref{ses2}), we can now express the functional in terms of $\overline{\xi}$, $\overline{\sigma}$ and $\beta$ writing $S=S[\rho_1]$ and  $E^{f.g.}=E^{f.g.}[\rho_1]$. After straightforward calculations, we obtain
\begin{equation}
S=-\frac{1}{2}\ln\beta,\label{ses4}
\end{equation}
\begin{equation}
E^{f.g.}=\frac{1}{2}\int \overline{\xi}\psi\, dydz+\frac{1}{2\beta}+\int \frac{\overline{\sigma}^2}{4y}\, dydz,
\label{ses5}
\end{equation}
\begin{equation}
H=\int \overline{\xi}\, \overline{\sigma}\, dydz,
\label{ses6}
\end{equation}
\begin{equation}
I=\int \overline{\sigma}\, dydz,
\label{ses7}
\end{equation}
up to some constant terms.  Note that $\beta$ is determined by the energy constraint  given by Eq. (\ref{ses5}) leading to
\begin{equation}
\frac{1}{2\beta}=E^{f.g.}-E^{c.g.}[\overline{\xi},\overline{\sigma}],
\label{ses8}
\end{equation}
where $E^{c.g.}$ is the macroscopic energy defined by Eq. (\ref{ecg}). This relation can be used to express the entropy  Eq. (\ref{ses4}) in terms of  $\overline{\xi}$ and $\overline{\sigma}$ alone.

(ii) {\it Second step:} we now have to solve the maximization problem
\begin{equation}
\max_{\overline{\sigma},\overline{\xi}}\lbrace S[\overline{\sigma},\overline{\xi}]\, | \, E^{f.g.}, \, H,\, I  \rbrace,\label{ses9}
\end{equation}
with
\begin{equation}
S=\frac{1}{2}\ln\left (2E^{f.g.}-\int \overline{\xi}\psi\, dydz-\int \frac{\overline{\sigma}^2}{2y}\, dydz\right ),
\label{ses10}
\end{equation}
\begin{equation}
H=\int \overline{\xi}\, \overline{\sigma}\, dydz,
\label{ses11}
\end{equation}
\begin{equation}
I=\int \overline{\sigma}\, dydz.
\label{ses12}
\end{equation}

(iii) {\it Conclusion:} Finally, the solution of (\ref{bes1})  is given by Eq. (\ref{ses2}) where $\overline{\sigma}$ is solution of (\ref{ses9}). Therefore, (\ref{bes1}) and (\ref{ses9}) are equivalent but (\ref{ses9}) is easier to solve because it is expressed in terms of  $\overline{\sigma}$ and $\overline{\xi}$ while (\ref{bes1}) is expressed in terms of $\rho$ and $\overline{\xi}$.

Up to second order, the variations of entropy given by Eq. (\ref{ses10}) are
\begin{eqnarray}
\Delta S=-\frac{1}{2}\beta\biggl (2\int \psi\delta\overline{\xi}\, dydz+\int \delta\psi\delta\overline{\xi}\, dydz\nonumber\\
+\int \frac{(\delta\overline{\sigma})^2}{2y}\, dydz+\int \frac{\overline{\sigma}\delta\overline{\sigma}}{y}\, dydz\biggr)\nonumber\\
-\beta^2\biggl (\int  \psi\delta\overline{\xi}\, dydz+\int\frac{\overline{\sigma}\delta\overline{\sigma}}{2y}\, dydz \biggr )^2,
\label{ses13}
\end{eqnarray}
where $\beta$ is given by Eq. (\ref{ses8}). The critical points of (\ref{ses9}) are determined by the variational problem
\begin{equation}
\delta S-\mu\delta H-\alpha\delta I=0.
\label{ses14}
\end{equation}
The variations on $\overline{\xi}$ yield
\begin{equation}
\beta {\psi} + \mu \overline{\sigma} = 0, \label{ses15}
\end{equation}
and the variations on $\overline{\sigma}$ yield
\begin{gather}
\overline{\sigma}=-\frac{2y}{\beta}( \mu \overline{\xi}+\alpha).\label{ses16}
\end{gather}
This returns the equations (\ref{bes8}), (\ref{bes10}) for the mean flow. Together with Eq. (\ref{ses2}), we recover the Gibbs state given by Eq. (\ref{bes9}). Considering now the second variations of entropy given by Eq. (\ref{ses13}), we find that a critical point of (\ref{ses9}) is a  maximum of $S$ at fixed microscopic energy, helicity and angular momentum iff
\begin{eqnarray}
-\frac{1}{2}\beta\biggl (\int \delta\psi\delta\overline{\xi}\, dydz
+\int \frac{(\delta\overline{\sigma})^2}{2y}\, dydz\biggr)
-\mu\int \delta\overline{\xi}\delta\overline{\sigma}\, dydz\nonumber\\
-\beta^2\biggl (\int  \psi\delta\overline{\xi}\, dydz+\int\frac{\overline{\sigma}\delta\overline{\sigma}}{2y}\, dydz \biggr )^2\le 0,\qquad
\label{ses17}
\end{eqnarray}
for all perturbations $\delta\overline{\xi}$ and  $\delta\overline{\sigma}$ that conserve helicity and angular momentum at first order (the conservation of microscopic energy has been  automatically taken into account in our formulation). The stability criterion (\ref{ses17}) is equivalent to Eq. (\ref{bes12}) but it is much simpler because it depends only on the perturbations $\delta\overline{\xi}$ and $\delta\overline{\sigma}$ instead of the perturbations  $\delta\rho$ of the full distribution of angular momentum. In fact, the stability condition (\ref{ses17}) can be further simplified. Indeed, using Eqs. (\ref{ses15}) and (\ref{ses16}), we find that the last term in parenthesis can be written
\begin{eqnarray}
\int  \psi\delta\overline{\xi}\, dydz+\int\frac{\overline{\sigma}\delta\overline{\sigma}}{2y}\, dydz=\nonumber\\
-\frac{\mu}{\beta}\int \left (\overline{\sigma}\delta\overline{\xi}+\overline{\xi}\delta\overline{\sigma}\right )\, dydz-\frac{\alpha}{\beta}\int\delta\overline{\sigma}\, dydz,
\end{eqnarray}
and it vanishes since the helicity and the angular momentum are conserved at first order so that $\delta H=\int \left (\overline{\sigma}\delta\overline{\xi}+\overline{\xi}\delta\overline{\sigma}\right )\, dydz=0$ and $\delta I=\int \delta\overline{\sigma}\, dydz=0$. Therefore, a critical point of (\ref{ses9}) is a  maximum of entropy at fixed microscopic energy, helicity and angular momentum iff
\begin{eqnarray}
-\frac{1}{2}\beta\biggl (\int \delta\psi\delta\overline{\xi}\, dydz
+\int \frac{(\delta\overline{\sigma})^2}{2y}\, dydz\biggr)\nonumber\\
-\mu\int \delta\overline{\xi}\delta\overline{\sigma}\, dydz\le 0,
\label{crit}
\end{eqnarray}
for all perturbations $\delta\overline{\xi}$ and  $\delta\overline{\sigma}$ that conserve helicity and angular momentum at first order. In fact, this stability condition can be obtained more rapidly if we remark that the maximization problem  (\ref{ses9}) is equivalent to the minimization of the macroscopic energy at fixed helicity and angular momentum (see Sec. \ref{sec_demin}).

\subsection{Equivalence with the minimum energy principle}
\label{sec_demin}

Since $\ln(x)$ is a monotonically increasing function, it is clear that the maximization problem (\ref{ses9}) is equivalent to
\begin{equation}
\min_{\overline{\sigma},\overline{\xi}}\lbrace E^{c.g.}[\overline{\sigma},\overline{\xi}]\, | \,  H,\, I  \rbrace,\label{ses9b}
\end{equation}
with
\begin{equation}
E^{c.g.}=\frac{1}{2}\int \overline{\xi}\psi\, dydz+\int \frac{\overline{\sigma}^2}{4y}\, dydz,
\label{ses10b}
\end{equation}
\begin{equation}
H=\int \overline{\xi}\, \overline{\sigma}\, dydz,
\label{ses11b}
\end{equation}
\begin{equation}
I=\int \overline{\sigma}\, dydz.
\label{ses12b}
\end{equation}
We have the equivalence
\begin{equation}
(\ref{ses9b}) \Leftrightarrow (\ref{ses9}) \Leftrightarrow (\ref{bes1}).\label{res3}
\end{equation}
Therefore, the maximization of entropy at fixed microscopic energy, helicity and angular momentum is equivalent to the minimization of macroscopic energy at fixed helicity and angular momentum.  The solution of (\ref{bes1}) is given by Eq. (\ref{ses2}) where $\overline{\sigma}$ is solution of (\ref{ses9b}). Therefore  (\ref{bes1}) and (\ref{ses9b}) are equivalent but (\ref{ses9b}) is easier to solve because it is expressed in terms of $\overline{\sigma}$ instead of $\rho$. Our approach therefore provides a justification of the minimum energy principle in terms of statistical mechanics. Note that, according to (\ref{gr1}), the principle (\ref{ses9b}) also assures that the mean flow associated with the statistical equilibrium state is nonlinearly dynamically stable with respect to the axisymmetric Euler equations.

The critical points of (\ref{ses9b}) are given by the variational problem
\begin{equation}
\delta E^{c.g.}+\mu\delta H+\alpha\delta I=0.\label{res15}
\end{equation}
The variations on $\overline{\xi}$ yield
\begin{equation}
 {\psi} + \mu \overline{\sigma} = 0, \label{res16}
\end{equation}
and the variations on $\overline{\sigma}$ yield
\begin{gather}
\overline{\sigma}=-2y( \mu \overline{\xi}+\alpha).\label{res17}
\end{gather}
This returns Eqs. (\ref{bes8}) and  (\ref{bes10}) for the mean flow (up to a trivial redefinition of $\mu$ and $\alpha$). Together with Eq. (\ref{ses2}), we recover the Gibbs state given by Eq. (\ref{bes9}).  On the other hand, this state is a minimum of $E^{c.g.}$ at fixed $H$ and $I$ iff
\begin{eqnarray}
\frac{1}{2}\int \delta\psi\delta\overline{\xi}\, dydz
+\int \frac{(\delta\overline{\sigma})^2}{4y}\, dydz\nonumber\\
+\mu\int \delta\overline{\xi}\delta\overline{\sigma}\, dydz\ge 0,\qquad\label{res18}
\end{eqnarray}
for all perturbations $\delta\overline{\xi}$ and  $\delta\overline{\sigma}$ that conserve helicity and angular momentum at first order. This is equivalent to the criterion given by Eq. (\ref{crit}) as it should.

We have thus shown the equivalence between the maximization of Boltzmann entropy at fixed helicity, angular momentum and fine-grained energy with the minimization of coarse-grained energy at fixed helicity and angular momentum. This equivalence has been shown here for global maximization. In Appendix \ref{sec_eqloc}, we prove the equivalence for local maximization by showing that the stability criteria (\ref{bes12}) and (\ref{res18}) are equivalent.

\subsection{Equivalence with the canonical ensemble}
\label{sec_eqcan}

The basic maximization problem (\ref{bes1}) is associated with the
microcanonical ensemble since the energy $E^{f.g.}$ is fixed. We could
also introduce a canonical ensemble where the inverse temperature
$\beta$ is fixed by making a Legendre transform $J=S-\beta E^{f.g.}$
of the entropy with respect to the energy \footnote{This can also be
viewed as a Legendre transform with respect to the {\it fragile
constraint} in the spirit of \cite{eht,physicaD,bouchet,proc}.}. The
corresponding maximization problem is
\begin{equation}
\max_{\rho,\overline{\xi}}\lbrace J[\rho]\, | \, H,\, I, \, \int \rho d\eta=1  \rbrace.\label{res2}
\end{equation}
A solution of (\ref{res2}) is always a solution of the more constrained
dual problem (\ref{bes1}) but the reciprocal is wrong in case of ensembles
inequivalence. In the present case, however,
we shall show that the microcanonical ensemble (\ref{bes1}) and the
canonical ensemble (\ref{res2}) are equivalent. This is
because the fluctuations of the energy are quadratic.

To solve the maximization problem (\ref{res2}) we can
proceed in two steps. We first maximize $J$ at fixed $H$, $I$, $\int
\rho\, d\eta=1$ {\it and} $\overline{\sigma}({\bf r})=\int\rho\eta\,
d\eta$ and $\overline{\xi}({\bf r})$. This is equivalent to maximizing
$\tilde{J}=S-\beta \int \rho \frac{\eta^2}{4y}\, dydzd\eta$ at fixed
$\int \rho\, d\eta=1$ {and} $\overline{\sigma}({\bf r})=\int\rho\eta\,
d\eta$. This leads to the optimal distribution (\ref{ses2}) where
$\beta$ is now fixed. This is clearly the global maximum of
$\tilde{J}$ with the previous constraints. Using this optimal
distribution, we can now express the free energy in terms of
$\overline{\xi}$ and $\overline{\sigma}$ by writing
$J[\overline{\xi},\overline{\sigma}]=J[\rho_1]$. After straightforward
calculations, we obtain
\begin{equation}
J=-\beta E^{c.g.},
\label{res9}
\end{equation}
up to some constant terms (recall that $\beta$ is a fixed parameter in the present ``canonical" situation). In the second step, we have to solve the maximization problem
\begin{equation}
\max_{\overline{\sigma},\overline{\xi}}\lbrace J[\overline{\sigma},\overline{\xi}]\, | \, H,\, I  \rbrace.
\label{res12}\end{equation}
Finally, the solution of (\ref{res2}) is given by (\ref{ses2}) where
$\overline{\sigma}$ is determined by (\ref{res12}). Therefore, the
canonical variational principle (\ref{res2}) is equivalent to
(\ref{res12}). On the other hand, since $\beta>0$, the maximization
problem (\ref{res12}) with Eq. (\ref{res9}) is equivalent to
(\ref{ses9b}). Since we have proven previously that (\ref{ses9b}) is
equivalent to the microcanonical variational principle (\ref{bes1}),
we conclude that the microcanonical and canonical ensembles are
equivalent.

\section{Conclusion} \label{conclusion}

In the present paper, we have constructed a simplified thermodynamic
approach of the axisymmetric Euler equations so as to describe its statistical equilibrium states. This predicts the mean field at metaequilibrium and the fluctuations around it.
We have considered two mean field theories.
In the first one, we have ignored the fluctuations of vorticity. In that case, we have found that the fluctuations of angular momentum are Gaussian and that the mean flow is in a Beltrami state. Furthermore, we have proven that the maximization of entropy at fixed  helicity, angular momentum and microscopic energy is equivalent to the minimization of macroscopic energy at fixed helicity and angular momentum. This provides a justification of this selective decay principle from statistical mechanics. These results are very similar to the case of 2D turbulence if we make the analogy between the angular momentum (axisymmetric) and the vorticity (2D). Indeed, in the simplified statistical approach of the 2D Euler equations developed in \cite{minens}, the fluctuations of vorticity are Gaussian and the mean flow is characterized by a linear $\overline{\omega}-\psi$ relationship. Furthermore, it has been proven that the maximization of entropy at fixed energy, circulation and microscopic enstrophy is equivalent to the minimization of macroscopic enstrophy at fixed energy and circulation. This provides a justification of the minimum enstrophy principle from statistical mechanics. In the second mean field theory, we have ignored the fluctuations of angular momentum. In that case, we have found again that the fluctuations of potential vorticity are Gaussian and that the mean flow is in a Beltrami state. We have also observed an interesting signature of vorticity stretching, via divergency of the variance of the vorticity fluctuations with increasing number of degrees of freedom.  Overall, the variance  of  fluctuations provides a measure of the number of degrees of freedom and of the statistical temperature(s) of turbulence, and allows to check Fluctuation-Dissipation Relations (FDR).

The question is whether these results are applicable to a laboratory
flow such as von K\'arm\'an flow. On the one hand, several basic
hypotheses are not satisfied in the VK flow: it is a
dissipative, forced flow, and instantaneous velocity fields are not
axisymmetric. On the other hand, one observes that in a stationary
state, dissipation and forcing balance globally, and may be neglected
locally, at least within an inertial range of scales, and the mean
flow is axisymmetric. This motivated experimental tests of the equilibrium and fluctuations relations, reported in \cite{monchaux2} in a
large Reynolds number von K\'arm\'an flow. It has been found that the observed stationary states are well described by the equilibrium states of the Euler equations, and that both the velocity and vorticity fluctuations are approximately uniform over the box. These fluctuations depend on three unknowns: the mean field statistical temperatures $1/\beta_\xi$ and $1/\beta_\sigma$, and the effective number of degrees of freedom $N_{tot}$. One can therefore use the experimental measurements to estimate these parameters under a supplementary assumption. Assuming $N_{tot}=1$, one finds that $\beta_\sigma$ and $\beta_\xi$ differ by an order of magnitude \cite{monchaux2}, an intriguing result that may come from the out-of-equilibrium character of the turbulence. On the other hand, assuming $\beta_\sigma=\beta_\xi$ one finds that $N_{tot}$ is of the order of $4$ to $6$, a rather small number for a turbulent flow. This may be due to strong correlations within the flow, that effectively reduces the number of degrees of freedom.

Altogether, these results are an indication that thermodynamics of
Euler axisymmetric flows can bring new interesting information about
real flows. Indeed, our statistical theory of axisymmetric flows that
can account for certain experimental results reported in
\cite{monchaux1,monchaux2}. In a forthcoming communication
\cite{naso2} we shall explain from this approach a turbulent
bifurcation that has been observed in a von K\'arm\'an flow
\cite{ravelet04}. Despite its good agreement with experiments, several
criticisms can be made to our approach. For example, in contrast with
what is hypothesized or derived in the paper, it is indeed usually
believed that the statistics of fluctuations in turbulence are
essentially non-Gaussian, that the mean-field approximation is not
satisfactory, and that the forcing and dissipation are important in
the process. We remark that these properties have been observed for
homogeneous and isotropic (three dimensional) turbulence. In our
$2.5$D situation, these general results may not be correct anymore,
since the dominant dynamical processes in both systems are probably
different, in particular because of the quasi-2D nature of
axisymmetric turbulence. In our case, it is not unlikely that the
distribution of angular momentum is Gaussian (or close to Gaussian),
like the distribution of the velocity components in 3D
turbulence. However, the distribution of azimuthal vorticity, which is
a derivative, may not be Gaussian because of
intermittency. Unfortunately, we are not able to check these
predictions experimentally due to a lack of statistics.

We have applied the maximum entropy principle to the Euler equation. Alternatively, Adzhemyan \& Nalimov \cite{an1,an2} have applied this principle to the stochastic Navier-Stokes equation. The renormalization group approach was used instead of the mean-field approximation, a non-Gaussian distribution was obtained and the Kolmogorov spectrum was derived for the inertial range. It would be interesting to extend their approach to our $2.5$ situation to see what it predicts and compare with our results.

This work was supported by European Contract WALLTURB.

\appendix

\section{Generalization and link with other results} \label{gc}

In the present paper, we have considered a restricted class of flows
for which the only invariants are $E$, $H=H_1$ and $I=I_1$. The
complete generalization to arbitrary invariants of axisymmetric Euler
equations ($E$, $H_F$, $I_G$) remains an unsolved problem. There are
however some special cases where we can perform the maximization
problem and find equilibrium distributions. We list these cases below,
and make connection with previous results.

\subsection{Conservation of $E^{c.g.}$, $H=H_1$, $\Gamma=H_0$, $I=I_1$ and $I_{n>1}^{f.g.}$ }
\label{sec_lep}

Leprovost {\it et al.} \cite{leprovost} have considered the maximization
problem
\begin{equation}
\max_{\rho,\overline{\xi}}\lbrace S[\rho]\, | \, E^{c.g.}, \, H,\, \Gamma,\, I, \, I_{n>1}^{f.g.},\, \int \rho d\eta=1  \rbrace,
\label{gc1}
\end{equation}
where $S[\rho]$ is the mixing entropy (\ref{mixing}) and $I_{n>1}^{f.g.}=\int \rho\eta^n\, dydzd\eta$. If we make a Legendre transform of the entropy with respect to the fragile constraints, we obtain the reduced maximization problem
\begin{equation}
\max_{\rho,\overline{\xi}}\lbrace S_{\chi}[\rho]\, | \, E^{c.g.}, \, H,\, \Gamma,\, I, \, \int \rho d\eta=1  \rbrace,
\label{gc2}
\end{equation}
with
\begin{equation}
S_{\chi}[\rho]=S[\rho]-\sum_{n>1}\alpha_nI_{n}^{f.g.}.
\label{gc2b}
\end{equation}
Explicitly
\begin{equation}
S_{\chi}[\rho]=-\int \rho\ln\left\lbrack\frac{\rho}{\chi(\eta)}\right\rbrack\, dydzd\eta,
\label{gc2bbis}
\end{equation}
where $\chi(\eta)=e^{-\sum_{n>1}\alpha_n\eta^n}$. Proceeding as in \cite{ceht}, we can show that the maximization problem  (\ref{gc2}) is equivalent to
\begin{equation}
\max_{\overline{\sigma},\overline{\xi}}\lbrace S[\overline{\sigma}]\, | \, E^{c.g.}, \, H,\, \Gamma,\, I   \rbrace,
\label{gc3}
\end{equation}
where $S[\overline{\sigma}]$ is the generalized entropy
\begin{equation}
S=-\int C(\overline{\sigma})\, d{\bf r},\qquad  C(\overline{\sigma})=-\int^{\overline{\sigma}}\lbrack(\ln\hat{\chi})'\rbrack^{-1}(-x)\, dx.
\label{gc4}
\end{equation}
The critical points of (\ref{gc3}) are given by
\begin{equation}
\beta\psi=-\mu\overline{\sigma}-\gamma,
\label{ph1}
\end{equation}
\begin{equation}
-C'(\overline{\sigma})=\beta\frac{\overline{\sigma}}{2y}+\mu\overline{\xi}+\alpha.
\label{ph2}
\end{equation}
The solutions of (\ref{gc2}) and (\ref{gc3}) are always solutions of
(\ref{gc1}) but the reciprocal is wrong in case of ensemble
inequivalence. Thus  $(\ref{gc2}) \Leftrightarrow  (\ref{gc3}) \Rightarrow (\ref{gc1})$. If we consider the particular case where $\alpha_n=0$ for $n\neq
2$ and $\alpha_2\neq 0$, then, proceeding as in \cite{ceht}, we find
that the fluctuations of angular momentum are Gaussian with variance $\sigma_2=1/(2\alpha_2)$ and that the
generalized entropy is 
\begin{equation}
S=-\frac{1}{2\sigma_2}\int \overline{\sigma}^2\, d{\bf r}=-\frac{1}{2\sigma_2}I_2^{c.g.}.
\label{gc4b}
\end{equation}
Since $\sigma_2>0$, the generalized entropy is proportional to minus $I_{2}^{c.g.}$. The corresponding critical points 
\begin{equation}
\beta\psi=-\mu\overline{\sigma}-\gamma,
\label{ph1add}
\end{equation}
\begin{equation}
-\frac{\overline{\sigma}}{\sigma_2}=\beta\frac{\overline{\sigma}}{2y}+\mu\overline{\xi}+\alpha,
\label{ph2add}
\end{equation}
are steady states of the axisymmetric Euler equations corresponding to $f(\psi)$ linear and $g(\psi)$ linear but not constant. Therefore, the statistical approach of Leprovost {\it et al.} \cite{leprovost} based on (\ref{gc1}) does {\it not} lead to Beltrami states (corresponding to $f$ linear and $g$ constant) contrary to the statistical approach developed in the present paper. 

Finally, if we consider the maximization problem
\begin{equation}
\max_{\rho,\overline{\xi}}\lbrace S[\rho]\, | \, E^{c.g.}, \, H,\, \Gamma,\, I, \, I_{2}^{f.g.},\, \int \rho d\eta=1  \rbrace,
\label{gc1az}
\end{equation} 
where only the quadratic integral $I_{2}^{f.g.}$ is conserved among the set of fragile constraints, and proceed as in \cite{minens}, we find that (\ref{gc1}) is equivalent to
\begin{equation}
\min_{\overline{\sigma},\overline{\xi}}\lbrace I_2^{c.g.}[\overline{\sigma}]\, | \, E^{c.g.}, \, H,\, \Gamma,\, I   \rbrace.
\label{gc3frs}
\end{equation}
Since  (\ref{gc3frs}) is equivalent to (\ref{gc3}) with  (\ref{gc4b}), hence to (\ref{gc2}), we also conclude that, in the specific case where only $I_{2}^{f.g.}$ is conserved among the set of fragile constraints, (\ref{gc1az}) is equivalent to (\ref{gc2}) with $\alpha_n=0$ for $n\neq 2$.

\subsection{Conservation of $E^{c.g.}$, $H=H_1$, $\Gamma=H_0$ and $I=I_1$ }

We consider the maximization problem
\begin{equation}
\max_{\rho,\overline{\xi}}\lbrace S[\rho]\, | \, E^{c.g.}, \, H,\,\Gamma,\, I, \, \int \rho d\eta=1  \rbrace,
\label{a1}
\end{equation}
The variations over $\overline{\xi}$ give
\begin{equation}
\beta\psi=-\mu\overline{\sigma}-\gamma,
\label{a2}
\end{equation}
and the variations over $\rho$ give the exponential distribution
\begin{equation}
\rho=\frac{1}{Z}e^{-\left (\beta\frac{\overline{\sigma}}{2y}+\mu\overline{\xi}+\alpha\right )\eta}.
\label{a3}
\end{equation}
Since this distribution is not normalizable, we must impose some bounds on the angular momentum and we shall assume $-\lambda<\sigma<\lambda$ (symmetric). In that case, we have
\begin{equation}
\overline{\sigma}=\lambda L\left \lbrack -\lambda\left (\beta\frac{\overline{\sigma}}{2y}+\mu\overline{\xi}+\alpha\right )\right \rbrack,
\label{a4}
\end{equation}
where
\begin{equation}
L(x)=\tan^{-1}(x)-\frac{1}{x},
\label{a5}
\end{equation}
is the Langevin function \cite{turkington}. We see that $(\overline{\xi},\overline{\sigma})$ is a steady state of the axisymmetric Euler equations. If we consider the maximization problem
\begin{equation}
\max_{\rho,\overline{\sigma}}\lbrace S[\rho]\, | \, E^{c.g.}, \, H,\,\Gamma,\, I, \, \int \rho d\nu=1  \rbrace,
\label{a1by}
\end{equation}
we find symmetric results but, in that case, $(\overline{\xi},\overline{\sigma})$ is {\it not} a steady state of the axisymmetric Euler equations.

\subsection{Conservation of $E^{f.g.}$, $H=H_1$ and $I=I_1$}

In the present paper, we have considered
the maximization problem
\begin{equation}
\max_{\rho,\overline{\xi}}\lbrace S[\rho]\, | \, E^{f.g.}, \, H,\, I, \, \int \rho d\eta=1  \rbrace.
\label{gc5}
\end{equation}
The variations over $\overline{\xi}$ give
\begin{equation}
\beta\psi+\mu\overline{\sigma}=0
\label{gc180}
\end{equation}
and the variations over $\rho$ give the Gaussian distribution
\begin{equation}
\rho=\frac{1}{Z}e^{-\frac{\beta\eta^2}{4y}-(\mu\overline{\xi}+\alpha)\eta}.
\label{gc190}
\end{equation}
We have
\begin{equation}
\overline{\sigma}=-\frac{2y}{\beta}\left(\mu\overline{\xi}+\alpha\right),\quad \sigma_2=\frac{2y}{\beta}.
\label{gc200}
\end{equation}
In that case $(\overline{\xi},\overline{\sigma})$ is a steady state of the
Euler equations with $f$ linear and $g$ constant (Beltrami state). The stream function can be expressed in terms of Bessel functions. Furthermore, we have shown that (\ref{gc5}) is equivalent to 
\begin{equation}
\min_{\overline{\sigma},\overline{\xi}}\lbrace E^{c.g.}[\overline{\sigma}]\, | \,  H,\, I   \rbrace.
\label{gc8}
\end{equation}

\subsection{General case: conservation of $E^{f.g.}$, $H=H_1$, $I$, $H_{n>1}^{f.g.}$, $I_{n>1}^{f.g.}$}

Let us consider the problem
\begin{equation}
\max_{\rho,\overline{\xi}}\lbrace S[\rho]\, | \, E^{f.g.}, \, H,\, I, \, H_{n>1}^{f.g.}, \, I_{n>1}^{f.g.}, \, \int \rho d\eta=1  \rbrace,
\label{gc9}
\end{equation}
generalizing the one studied in the present paper. The variations over $\overline{\xi}$ give
\begin{equation}
\beta\psi+\mu\overline{\sigma}+\sum_{n>1}\mu_n\overline{\sigma^n}=0,
\label{gc10}
\end{equation}
and the variations over $\rho$ give
\begin{equation}
\rho=\frac{1}{Z}e^{-\sum_{n>1}\alpha_n\eta^n}e^{-\sum_{n>1}
\mu_n\overline{\xi}\eta^n}e^{-\frac{\beta\eta^2}{4y}}e^{-(\mu\overline{\xi}+\alpha)\eta}.
\label{gc11}
\end{equation}
However, it is difficult to be more explicit. Therefore, we shall consider simpler problems.

\subsection{Conservation of $E^{f.g.}$, $H=H_1$, $I=I_1$, and $I_{2}^{f.g.}$}

We consider the maximization problem
\begin{equation}
\max_{\rho,\overline{\xi}}\lbrace S[\rho]\, | \, E^{f.g.}, \, H,\, I, \, I_{2}^{f.g.}, \, \int \rho d\eta=1  \rbrace.
\label{gc17}
\end{equation}
The variations over $\overline{\xi}$ give
\begin{equation}
\beta\psi+\mu\overline{\sigma}=0,
\label{gc18}
\end{equation}
and the variations over $\rho$ give the Gaussian distribution
\begin{equation}
\rho=\frac{1}{\sqrt{2\pi\sigma_2}}e^{-\frac{(\eta-\overline{\sigma})^2}{2\sigma_2}},
\label{gc19}
\end{equation}
with
\begin{equation}
\overline{\sigma}=-\frac{\mu\overline{\xi}+\alpha}{2(\alpha_2+\frac{\beta}{4y})},\quad
\sigma_2=\frac{1}{2(\alpha_2+\frac{\beta}{4y})}.
\label{gc20}
\end{equation}
In that case $(\overline{\xi},\overline{\sigma})$ is a steady state of the
Euler equations with $f$ and $g$ linear.  The stream function can be expressed in terms of Whittaker functions \cite{leprovost}. If we make a Legendre transform of the entropy with respect to the fragile constraints, we obtain the reduced maximization problem
\begin{equation}
\max_{\rho,\overline{\xi}}\lbrace {\cal S}[\rho]\, | \, H,\, I, \, \int \rho d\eta=1  \rbrace,
\label{ra1}
\end{equation} 
with
\begin{equation}
{\cal S}=S-\beta E^{f.g.}-\alpha_2 I_2^{f.g.}.
\label{ra2}
\end{equation} 
We proceed as in \cite{ceht}. We first maximize ${\cal S}$ at fixed $H$, $I$, normalization {\it and} $\overline{\sigma}=\int \rho\eta\, d\eta$. This yields an optimal density $\rho_1({\bf r},\eta)$ given by Eq. (\ref{gc19}) where $\sigma_2$ is given by Eq. (\ref{gc20})-b. Then, we find that the maximization problem (\ref{ra1}) is equivalent to
\begin{equation}
\max_{\overline{\sigma},\overline{\xi}}\lbrace {S}[\overline{\sigma},\overline{\xi}]\, | \, H,\, I\rbrace,
\label{ra3}
\end{equation}  
with the generalized entropy  ${S}[\overline{\sigma},\overline{\xi}]\equiv {\cal S}[\rho_1]$. An explicit calculation leads to
\begin{equation}
{S}[\overline{\sigma},\overline{\xi}]=-\beta E^{c.g.}-\alpha_{2} I_{2}^{c.g.}.
\label{ra4}
\end{equation}  
This is a sort of ``mixed'' case between the one studied in the main part of the paper (leading to the minimization of $E^{c.g.}$) and the one discussed at the beginning of this Appendix (leading to the minimization of $I_2^{c.g.}$).
The critical points of (\ref{ra3}) return Eqs. (\ref{gc18}) and (\ref{gc20})-a. Furthermore, a solution of (\ref{ra1}) or (\ref{ra3}) is always a solution of (\ref{gc17}) but the reciprocal is wrong in case of ensemble inequivalence. We have $(\ref{ra1}) \Leftrightarrow  (\ref{ra3}) \Rightarrow  (\ref{gc17})$.

\subsection{Conservation of $E^{f.g.}$, $H=H_1$, $I=I_1$, $H_{2}^{f.g.}$ and $I_{2}^{f.g.}$}

We consider the maximization problem
\begin{equation}
\max_{\rho,\overline{\xi}}\lbrace S[\rho]\, | \, E^{f.g.}, \, H,\, I, \, H_{2}^{f.g.}, \, I_{2}^{f.g.}, \, \int \rho d\eta=1  \rbrace,
\label{gc12}
\end{equation}
where $S[\rho]$ is the mixing entropy (\ref{mixing}) and $H_{n>1}^{f.g.}=\int \overline{\xi}\rho\eta^n\, dydzd\eta$. The variations over $\overline{\xi}$ give
\begin{equation}
\beta\psi+\mu\overline{\sigma}+\mu_2\overline{\sigma^2}=0
\label{gc13}
\end{equation}
and the variations over $\rho$ give the Gaussian distribution
\begin{equation}
\rho=\frac{1}{Z}e^{-\alpha_2\eta^2}e^{-\mu_2\overline{\xi}\eta^2}e^{-\frac{\beta\eta^2}{4y}}e^{-(\mu\overline{\xi}+\alpha)\eta}.
\label{gc14}
\end{equation}
We have
\begin{equation}
\overline{\sigma}=-\frac{\mu\overline{\xi}+\alpha}{2(\alpha_2+\mu_2\overline{\xi}+\frac{\beta}{4y})},
\label{gc15}
\end{equation}
\begin{equation}
\sigma_2=\frac{1}{2(\alpha_2+\mu_2\overline{\xi}+\frac{\beta}{4y})}.
\label{gc16}
\end{equation}

\section{Detailed proof of inequality (\ref{bel7})}
\label{A1}

We consider the minimization problem
\begin{eqnarray}
\label{stab1}
\min_{\xi,\sigma} \lbrace \, E[\xi,\sigma]\, | \, H, \, I \, \rbrace,
\end{eqnarray}
with
\begin{eqnarray}
\label{stab2}
E=\frac{1}{2}\int \xi\psi\, dydz+\frac{1}{4}\int \frac{\sigma^2}{y}\, dydz,
\end{eqnarray}
\begin{eqnarray}
\label{stab3}
H=\int \xi\sigma\, dydz,
\end{eqnarray}
\begin{eqnarray}
\label{stab4}
I=\int \sigma\, dydz.
\end{eqnarray}
We shall look for (local) minima of energy at fixed helicity and angular momentum. We proceed as in \cite{proc,cc}. The variations of these functionals up to second order are
\begin{eqnarray}
\label{stab5}
\Delta E=\int \psi\delta\xi\, dydz+\frac{1}{2}\int \delta\xi\delta\psi\, dydz\nonumber\\
+\frac{1}{2}\int \frac{\sigma\delta\sigma}{y}\, dydz+\frac{1}{4}\int \frac{(\delta\sigma)^2}{y}\, dydz,
\end{eqnarray}
\begin{eqnarray}
\label{stab6}
\Delta H=\int \xi\delta\sigma\, dydz+\int \sigma\delta\xi\, dydz+\int\delta\xi\delta\sigma\, dydz,
\end{eqnarray}
\begin{eqnarray}
\label{stab7}
\Delta I=\int \delta\sigma\, dydz.
\end{eqnarray}
The critical points satisfy the variational principle for the first variations
\begin{eqnarray}
\label{stab8}
\delta E+\mu \delta H +\alpha \delta I=0.
\end{eqnarray}
Taking the variations over $\xi$ and $\sigma$, we obtain
\begin{gather}
\psi+\mu\sigma=0, \label{stab9} \\
\frac{\sigma}{2y}+\mu\xi+\alpha=0.\label{stab9b}
\end{gather}
A minimum of energy corresponds to $\Delta E>0$. Inserting Eqs.  (\ref{stab9}) and  (\ref{stab9b}) in Eq. (\ref{stab5}), we find that
\begin{eqnarray}
\label{stab10}
\Delta E=-\mu\int \sigma\delta\xi\, dydz+\frac{1}{2}\int \delta\xi\delta\psi\, dydz\nonumber\\
-\mu\int \xi\delta\sigma\, dydz-\alpha\int \delta\sigma\, dydz+\frac{1}{4}\int \frac{(\delta\sigma)^2}{y}\, dydz.\nonumber\\
\end{eqnarray}
Then, using Eqs. (\ref{stab6}) and (\ref{stab7}) with $\Delta H=\Delta I=0$, we obtain
\begin{eqnarray}
\Delta E=\frac{1}{2}\int\delta\xi\delta\psi\, dydz+\int \frac{(\delta\sigma)^2}{4y}\, dydz
+\mu\int \delta\xi\delta\sigma\, dydz. \nonumber\\
\label{stab11}
\end{eqnarray}
Therefore, a critical point of (\ref{stab1}) is a (local) minimum of energy at fixed helicity and angular momentum iff
\begin{eqnarray}
\frac{1}{2}\int\delta\xi\delta\psi\, dydz+\int \frac{(\delta\sigma)^2}{4y}\, dydz
+\mu\int \delta\xi\delta\sigma\, dydz\ge  0, \nonumber\\
\label{stab12}
\end{eqnarray}
for all perturbations $\delta\sigma$ and $\delta\xi$ that conserve helicity and angular momentum at first order. This amounts to having $\delta^2(E+\mu H+\alpha I)>0$ for all the perturbations that conserve $H$ and $I$ at first order.

\section{Equivalence between (\ref{bes12}) and (\ref{res18})}
\label{sec_eqloc}

In Sec. \ref{SectionConnection}, we have shown the equivalence of
(\ref{bes1}) and (\ref{ses9b}) for global maximization.  In this
Appendix, we show the equivalence of (\ref{bes1}) and (\ref{ses9b})
for local maximization, {\it i.e.} $\rho({\bf r},\eta)$ is a (local) maximum
of $S[\rho]$ at fixed $E^{f.g.}$, $H$, $I$ and normalization if, and
only if, the corresponding coarse-grained distribution of angular
momentum $\overline{\sigma}({\bf r})$ is a (local) minimum of
$E^{c.g.}[\overline{\sigma},\overline{\xi}]$ at fixed $H$ and $I$. To
that purpose, we show the equivalence between the stability criteria
(\ref{bes12}) and (\ref{res18}). We use a general method similar to the
one used in \cite{ceht,nyquistgrav,cc,minens} in related problems.

We shall determine the optimal perturbation $\delta\rho_{*}({\bf
r},\eta)$ that maximizes $\delta^{2}J[\delta\rho]$ given by
Eq. (\ref{bes12}) with the constraints $\delta\overline{\sigma}=\int
\delta\rho\eta\, d\eta$, $\delta E^{f.g.}=\int\psi\delta\overline{\xi}\, dydz+\int
\delta\rho \frac{\eta^2}{4y}\, d\eta dydz=0$ and $\int \delta\rho\,
d\eta=0$, where $\delta\overline{\sigma}({\bf r})$ and
$\delta\overline{\xi}({\bf r})$ are prescribed (they are only ascribed
to conserve $H$ and $I$ at first order). Since the specification of
$\delta\overline{\sigma}$ and $\delta\overline{\xi}$ (hence
$\delta\psi$) determine the second and third integrals in
Eq. (\ref{bes12}), we can write the variational problem in the form
\begin{eqnarray}
\label{eqa1}
\delta\left (-\frac{1}{2}\int\frac{(\delta\rho)^2}{\rho}\, dydzd\eta\right )-\int\lambda({\bf r})\delta\left (\int\delta\rho\eta\, d\eta\right )\, dydz\nonumber\\
-{\tilde\mu}\delta\left (\int \delta\rho\frac{\eta^2}{4y}\,  dydzd\eta\right ) - \int\zeta({\bf r})\delta\left (\int\delta\rho\, d\eta\right )\, dydz=0,\nonumber\\
\end{eqnarray}
where $\lambda({\bf r})$, ${\tilde\mu}$  and $\zeta({\bf r})$ are Lagrange multipliers. This gives
\begin{eqnarray}
\label{eqa2}
\delta\rho_*({\bf r},\eta)=-\rho({\bf r},\eta)\left \lbrack {\tilde\mu}\frac{\eta^2}{4y}+\lambda({\bf r})\eta+\zeta({\bf r})\right \rbrack,
\end{eqnarray}
and it is a global maximum of $\delta^{2}J[\delta\rho]$ with the previous constraints since $\delta^2(\delta^2 J)=-\int \frac{(\delta(\delta\rho))^2}{2\rho}\, dydzd\eta< 0$ (the constraints are linear in $\delta\rho$ so their second variations vanish). The Lagrange multipliers are determined from the above-mentioned constraints. The constraints $\int \delta\rho\, d\eta=0$ and $\delta\overline{\sigma}=\int \delta\rho\eta\, d\eta$ lead to
\begin{eqnarray}
\zeta({\bf r})+\lambda({\bf r})\overline{\sigma}({\bf r})+{\tilde\mu}\frac{\overline{\sigma^2}({\bf r})}{4y}=0,
\label{co1}
\end{eqnarray}
\begin{eqnarray}
\zeta({\bf r})\overline{\sigma}({\bf r})+\lambda({\bf r})\overline{\sigma^2}({\bf r})+{\tilde\mu}\frac{\overline{\sigma^3}({\bf r})}{4y}=-\delta\overline{\sigma}({\bf r}).
\label{co2}
\end{eqnarray}
Now, the state $\rho({\bf r},\eta)$ corresponds to the gaussian
distribution (\ref{bes9}). Therefore, we have the well-known relations
$\overline{\sigma^2}({\bf r})=\overline{\sigma}^2({\bf r})+\sigma_2$
and $\overline{\sigma^3}({\bf r})=\overline{\sigma}^3({\bf
r})+3\overline{\sigma}({\bf r})\sigma_2$ where
$\sigma_2=2y/\beta$. Substituting these relations in Eqs. (\ref{co1})
and (\ref{co2}), and solving for $\lambda({\bf r})$ and $\zeta({\bf
r})$, we obtain
\begin{eqnarray}
\lambda({\bf r})=-\frac{\beta}{2y}\delta\overline{\sigma}({\bf r})-\frac{{\tilde\mu}}{2y}\overline{\sigma}({\bf r}),
\end{eqnarray}
\begin{eqnarray}
\zeta({\bf r})=\frac{\beta}{2y}\overline{\sigma}({\bf r})\delta\overline{\sigma}({\bf r})+\frac{\tilde\mu}{4y}\overline{\sigma}^2({\bf r})-\frac{\tilde\mu}{2\beta}.
\end{eqnarray}
Therefore, the optimal perturbation (\ref{eqa2}) can be rewritten
\begin{eqnarray}
\label{optb}
\delta\rho_{*}=-\rho\left\lbrack -\frac{\beta}{2y}\delta\overline{\sigma}(\eta-\overline{\sigma})+{\tilde\mu}\left\lbrace \frac{1}{4y}(\eta-\overline{\sigma})^2-\frac{1}{2\beta}\right\rbrace\right\rbrack.\nonumber\\
\end{eqnarray}
The Lagrange multiplier ${\tilde\mu}$ is determined by substituting this expression in the constraint $\int\psi\delta\overline{\xi}\, dydz+\int\delta\rho\frac{\eta^2}{4y}\, dydzd\eta=0$. Using the well-known identity $\overline{\sigma^4}({\bf r})=\overline{\sigma}^4({\bf r})+6\sigma_2\overline{\sigma}^2({\bf r})+3\sigma_2^2$  valid for a gaussian distribution, we obtain after some simplifications
\begin{eqnarray}
\label{mu}
{\tilde\mu}=2\beta^2\left (\int\psi\delta\overline{\xi}\, dydz+\int \frac{\overline{\sigma}}{2y}\delta\overline{\sigma}\, dydz\right ).
\end{eqnarray}
Therefore, the optimal perturbation (\ref{eqa2}) is given by
Eq. (\ref{optb}) with Eq. (\ref{mu}). Since this perturbation
maximizes $\delta^{2}J[\delta\rho]$ with the above-mentioned
constraints, we have $\delta^{2}J[\delta\rho]\le
\delta^{2}J[\delta\rho_*]$. Explicating $\delta^{2}J[\delta\rho_*]$
using Eqs. (\ref{optb}) and (\ref{mu}), we obtain after simple
calculations
\begin{eqnarray}
\label{eqa3b}
\delta^{2}J[\delta\rho]\le -\frac{1}{2}\beta\biggl (\int \delta\psi\delta\overline{\xi}\, dydz
+\int \frac{(\delta\overline{\sigma})^2}{2y}\, dydz\biggr)\nonumber\\
-\mu\int \delta\overline{\xi}\delta\overline{\sigma}\, dydz
-\beta^2\biggl (\int  \psi\delta\overline{\xi}\, dydz+\int\frac{\overline{\sigma}\delta\overline{\sigma}}{2y}\, dydz \biggr )^2.\nonumber\\
\end{eqnarray}
The r.h.s. returns the functional appearing in Eq. (\ref{ses17}). We
have already explained in Sec. \ref{simplerprin} that for the class of
perturbations that we consider ($\delta H=\delta I=0$) the last term
in parenthesis vanishes. Therefore, the foregoing inequality can be
rewritten
\begin{eqnarray}
\label{eqa4}
\delta^{2}J[\delta\rho]\le -\frac{1}{2}\beta\biggl (\int \delta\psi\delta\overline{\xi}\, dydz
+\int \frac{(\delta\overline{\sigma})^2}{2y}\, dydz\biggr)\nonumber\\
-\mu\int \delta\overline{\xi}\delta\overline{\sigma}\, dydz,\qquad\qquad
\end{eqnarray}
where the r.h.s. is precisely the functional appearing in
Eq. (\ref{res18}). Furthermore, there is equality in Eq. (\ref{eqa4})
iff $\delta\rho=\delta\rho_*$. This proves that the stability criteria
(\ref{bes12}) and (\ref{res18}) are equivalent. Indeed: (i) if
inequality (\ref{res18}) is fulfilled for all perturbations
$\delta\overline{\sigma}$ and $\delta\overline{\xi}$ that conserve
helicity and angular momentum at first order, then according to
Eq. (\ref{eqa4}), we know that inequality (\ref{bes12}) is fulfilled
for all perturbations $\delta\rho$ and $\delta\overline{\xi}$ that
conserve helicity, angular momentum, fine-grained energy and
normalization at first order; (ii) if there exists a perturbation
$\delta\overline{\sigma}_*$ that violates inequality (\ref{res18}),
then the perturbation $\delta\rho_*$ given by Eq. (\ref{optb}) with
Eq. (\ref{mu}) and $\delta\overline{\sigma}=\delta\overline{\sigma}_*$
violates (\ref{bes12}). In conclusion, the stability
criteria (\ref{bes12}) and (\ref{res18}) are equivalent.

\section{Relaxation equations}
\label{sec_rela}

\subsection{Relaxation equations associated with the maximization problem (\ref{bes1})}

Like in classical statistical physics, it may be interesting to derive
relaxation equations towards the equilibrium states so as to be able
to describe dynamical, non-stationary, regimes. On a practical point
of view, these relaxation equations can also provide a useful
numerical algorithm to solve the maximization problem (\ref{bes1}) and
be sure that we select entropy maxima (not minima or saddle
points). We follow the methodology described in \cite{proc}.  We
introduce a current of probability in the space of angular momentum
fluctuations $\eta$ and construct a set of relaxation equations that
increase $S[\rho]$ while conserving ${E}^{f.g.}$, ${I}$ and ${H}$
using a Maximum Entropy Production Principle (this can be viewed as
the variational formulation of Onsager's linear thermodynamics). The
dynamical equations that we consider can be written as
\begin{gather}
\frac{\partial\overline{\xi}}{\partial t}+{\bf u}\cdot \nabla\overline{\xi}=\frac{\partial}{\partial z} \left( \frac{\overline{\sigma^2}}{4y^2} \right)+X, \label{ber1} \\
\frac{\partial\rho}{\partial t}+{\bf u}\cdot \nabla\rho=-\frac{\partial J}{\partial \eta},\label{ber2}
\end{gather}
where $X$ and $J$ are two unknown quantities to be chosen so as to increase $S[\rho]$ while conserving ${E}^{f.g.}$, ${H}$ and ${I}$.
In the second equation, the local normalization $\int \rho d\eta=1$ is satisfied provided that $J \rightarrow 0$ as $\eta \rightarrow \pm \infty$. Multiplying Eq. (\ref{ber2})  by $\eta$ and integrating on all the levels, we get
\begin{equation}
\frac{\partial \overline{\sigma}}{\partial t}+{\bf u}\cdot \nabla\overline{\sigma}=\int J d\eta\equiv Y. \label{ber3}
\end{equation}
Next, multiplying Eq. (\ref{ber2})  by $\eta^2$ and integrating on all the levels, we obtain
\begin{equation}
\frac{\partial \overline{\sigma^2}}{\partial t}+{\bf u}\cdot \nabla\overline{\sigma^2}=2\int J\eta d\eta. \label{ber4}
\end{equation}
From Eqs. (\ref{ber3}) and (\ref{ber4}), we find that
\begin{equation}
\frac{\partial {\sigma_2}}{\partial t}+{\bf u}\cdot \nabla {\sigma_2}=2\int J(\eta-\overline{\sigma}) d\eta. \label{ber5}
\end{equation}
The time variations of $S[\rho]$ are given by
\begin{equation}
\dot{S}=-\int \frac{J}{\rho} \frac{\partial\rho}{\partial \eta}\, dydzd\eta,\label{ber6}
\end{equation}
while those of the invariants are given by
\begin{gather}
\dot{{E}}^{f.g.}=0=\int X {\psi} dydz + \int J\frac{\eta}{2y} dydzd\eta, \label{ber7}\\
\dot{{H}}=0=\int X \overline{\sigma} dydz+ \int J \overline{\xi} dydzd\eta,\label{ber8}\\
\dot{{I}}=0=\int J dydzd\eta.\label{ber8b}
\end{gather}
Following the Maximum Entropy Production Principle, we maximize $\dot{S}$ with $\dot{{E}}^{f.g.}=\dot{{H}}=\dot{{I}}=0$ and the additional constraints
\begin{equation}
\frac{X^2}{2}\le C_\xi({\bf r},t), \quad \int \frac{J^2}{2\rho} d\eta\le C({\bf r},t).\label{ber9}
\end{equation}
The variational principle can be written in the form
\begin{eqnarray}
\delta \dot{S} - \beta(t) \delta \dot{{E}}^{f.g.} - \mu(t) \delta \dot{{H}}- \alpha(t) \delta \dot{{I}}   \nonumber \\
-\int \frac{1}{D({\bf r},t)} \delta \left( \int \frac{J^2}{2\rho} d\eta \right) dydz  \nonumber \\
-\int \frac{1}{\chi({\bf r},t)} \delta \left( \frac{X^2}{2} \right) dydz &=& 0,\label{ber10}
\end{eqnarray}
where $\beta(t)$, $\mu(t)$, $\alpha(t)$, $D({\bf r},t)$ and $\chi({\bf r},t)$ are time dependent Lagrange multipliers associated with the constraints. This leads to the following optimal quantities
\begin{gather}
J=-D\left\lbrack \frac{\partial\rho}{\partial \eta}+\rho\left (\frac{\beta(t) \eta}{2y} + \mu(t) \overline{\xi}+\alpha(t) \right )\right\rbrack, \label{ber11} \\
X=-\chi \left( \beta(t){\psi} +\mu(t)\overline{\sigma} \right). \label{ber12}
\end{gather}
Therefore, the relaxation equation for the distribution of angular momentum is
\begin{eqnarray}
\frac{\partial\rho}{\partial t} &+&{\bf u}\cdot \nabla\rho\nonumber\\
&=&\frac{\partial}{\partial \eta} \left\lbrace D\left\lbrack \frac{\partial\rho}{\partial \eta}+\rho\left (\frac{\beta(t) \eta}{2y} + \mu(t) \overline{\xi}+\alpha(t) \right )\right\rbrack\right\rbrace.\nonumber\\\label{ber18}
\end{eqnarray}
Integrating Eq. (\ref{ber11}) on $\eta$ we get
\begin{equation}
Y=-D \left ( \frac{\beta(t)\overline{\sigma}}{2y} +\mu(t) \overline{\xi}+\alpha(t) \: \right ).\label{ber13}
\end{equation}
Inserting expressions (\ref{ber12}) and (\ref{ber13}) into Eqs.~(\ref{ber1}) and (\ref{ber3}) leads to the following relaxation equations for the mean flow
\begin{gather}
\frac{\partial \overline{\xi}}{\partial t}+{\bf u}\cdot \nabla\overline{\xi}=\frac{\partial}{\partial z} \left( \frac{\overline{\sigma^2}}{4y^2} \right) -\chi (\beta(t) {\psi}+ \mu(t) \overline{\sigma}),
 \label{ber19} \\
\frac{\partial \overline{\sigma}}{\partial t}+{\bf u}\cdot \nabla\overline{\sigma}=-D \left ( \frac{\beta(t)\overline{\sigma}}{2y} +\mu(t) \overline{\xi}+\alpha(t) \: \right ).\label{ber20}
\end{gather}
A relaxation equation can also be written for the centered variance $\sigma_2$. Using Eqs. (\ref{ber5}) and (\ref{ber11}), we obtain
\begin{equation}
\frac{\partial \sigma_2}{\partial t}+{\bf u}\cdot \nabla \sigma_2=2D \left (1  -  \frac{\beta(t) \sigma_2}{2y} \right). \label{ber21}
\end{equation}
Equations (\ref{ber20}) and ({\ref{ber21}) can be used to evaluate the evolution of $\overline{\sigma^2}=\overline{\sigma}^2+\sigma_2$.
The Lagrange multipliers evolve in time so as to satisfy the constraints. Substituting Eqs. (\ref{ber11}) and (\ref{ber12}) in Eqs. (\ref{ber7}), (\ref{ber8}) and (\ref{ber8b}), we obtain the algebraic equations
\begin{eqnarray}
\left (\left\langle\chi\psi^2\right\rangle + \left\langle D\frac{\overline{\sigma^2}}{4y^2}\right\rangle\right )\beta(t)+\left (\left\langle\chi\overline{\sigma}\psi\right\rangle
+\left\langle D\frac{\overline{\xi}\overline{\sigma}}{2y}\right\rangle\right )\mu(t)\nonumber\\
 + \left\langle D\frac{\overline{\sigma}}{2y}\right\rangle\alpha(t)
 =\left\langle D\frac{1}{2y}\right\rangle,\qquad\qquad\label{ber14}
\end{eqnarray}
\begin{eqnarray}
\left (\left\langle\chi\psi\overline{\sigma}\right\rangle + \left\langle D\frac{\overline{\sigma}\overline{\xi}}{2y}\right\rangle\right )\beta(t)+\left (\left\langle\chi\overline{\sigma}^2\right\rangle
+\left\langle D{\overline{\xi}^2}\right\rangle\right )\mu(t)\nonumber\\
 + \left\langle D\overline{\xi}\right\rangle\alpha(t)
 =0,\qquad\label{ber15}
\end{eqnarray}
\begin{eqnarray}
\left\langle D\frac{\overline{\sigma}}{2y}\right\rangle\beta(t)+\left\langle D\overline{\xi}\right\rangle\mu(t)+\alpha(t)\langle D\rangle
 =0.\qquad\label{ber16}
\end{eqnarray}
The coefficients $D$ and $\chi$, which can depend on $y$ and $z$, are not determined by the MEPP. They can be chosen so as to forbid divergency of the first term in the r.h.s. of equation (\ref{ber14}).

Substituting $\partial\rho/\partial\eta$ taken from Eq. (\ref{ber11}) in Eq. (\ref{ber6}) and using the constraints (\ref{ber7})-(\ref{ber8b}), we easily obtain
 \begin{eqnarray}
\dot S=\int \frac{J^2}{D\rho}\, dydzd\eta+\int \frac{X^2}{\chi}\, dydz,\label{ber17}
\end{eqnarray}
so that $\dot S\ge 0$ provided that $D$ and $\chi$ are both positive. On the other hand $\dot S=0$ iff $J=X=0$ leading to  the conditions of equilibrium  (\ref{bes8}) and (\ref{bes9}). From Lyapunov's direct method, we conclude that these relaxation equations tend to a maximum of entropy at fixed microscopic energy, helicity and angular momentum. Note that during the relaxation process, the distribution of angular momentum is not Gaussian but changes with time according to Eq. (\ref{ber18}).
The distribution is Gaussian only at equilibrium. Therefore, these relaxation equations describe not only the evolution of the mean flow but also the evolution of the distribution of fluctuations. We stress, however, that these equations are purely phenomenological and that there is no compelling reason why they should give an accurate description of the real dynamics. However, they can be used as a numerical algorithm to compute the equilibrium state corresponding to (\ref{bes1}). Indeed, these equations can only relax towards an entropy maximum at fixed microscopic energy, helicity and angular momentum, not towards a minimum or a saddle point that are linearly unstable with respect to these equations.

{\it Remark:} In fact, we will find in \cite{naso2} that there is no entropy maximum, just saddle points. In that case, the dynamical equations lead to a ``collapse'' at smaller and smaller scales, similar to the Richardson energy cascade in 3D turbulence. However, we will also observe that the system can remain blocked in a large-scale coherent structure (like in 2D turbulence). In the present 2.5D situation, this is an unstable state (saddle point of entropy), but it can persist for a long time if the dynamics does not spontaneously develop the ``dangerous'' perturbations that destabilize it. This is because a saddle point is unstable only for some perturbations but not for any perturbation.

\subsection{Relaxation equations associated with the maximization problem (\ref{ses9})}

We shall now introduce a set of relaxation equations associated with the maximization problem (\ref{ses9}). We write the dynamical equations as \footnote{In the present situation, $\sigma_2$ is given at each time by Eq. (\ref{ses3}) where $\beta(t)$ is given by Eq. (\ref{ser4}). Since $\sigma_2/y^2$ does not depend on $z$, we have written $\overline{\sigma}^2$ instead of $\overline{\sigma^2}$ in Eq. (\ref{ser1}).}:
\begin{gather}
\frac{\partial \overline{\xi}}{\partial t}+{\bf u}\cdot \nabla\overline{\xi}=\frac{\partial}{\partial z} \left( \frac{\overline{\sigma}^2}{4y^2} \right)+X,  \label{ser1}\\
\frac{\partial\overline{\sigma}}{\partial t}+{\bf u}\cdot \nabla\overline{\sigma}=Y,\label{ser2}
\end{gather}
where $X$ and $Y$ are two unknown quantities, to be chosen so as to increase $S[\overline{\xi},\overline{\sigma}]$ while conserving $E^{f.g.}$, $H$ and $I$ given by Eqs. (\ref{ses10}), (\ref{ses11}) and (\ref{ses12}).
The time variations of $S$ are
\begin{equation}
\dot{S}=-\frac{1}{2}\beta(t) \left (2\int \psi X\, dydz+\int \frac{\overline{\sigma}}{y}Y\, dydz\right ),\label{ser3}
\end{equation}
where $\beta(t)$ is determined by the constraint on the microscopic energy leading to
\begin{equation}
\frac{1}{\beta(t)}=2E^{f.g.}-\int \overline{\xi}\psi\, dydz-\int \frac{\overline{\sigma}^2}{2y}\, dydz.\label{ser4}
\end{equation}
On the other hand, the time variations of $H$ and $I$ are
\begin{gather}
\dot{{H}}=0=\int X \overline{\sigma} \, dydz+ \int Y\overline{\xi} \, dydz,\label{ser5}\\
\dot{{I}}=0=\int Y dydz.\label{ser6}
\end{gather}

Following the Maximum Entropy Production Principle, we maximize $\dot{S}$ with $\dot{{I}}=\dot{{H}}=0$ (the conservation of the microscopic energy has been taken into account in Eq. (\ref{ser4})) and the additional constraints
\begin{equation}
 \frac{X^2}{2}\le C_\xi({\bf r},t),\quad \frac{Y^2}{2}\le C_{\sigma}({\bf r},t).\label{ser7}
\end{equation}
The variational principle can be written in the form
\begin{eqnarray}
\delta \dot{S} - \mu(t) \delta \dot{{H}}- \alpha(t) \delta \dot{{I}}   \nonumber \\
-\int \frac{1}{\chi({\bf r},t)} \delta \left( \frac{X^2}{2} \right) dydz \nonumber \\
-\int \frac{1}{D({\bf r},t)} \delta \left(\frac{Y^2}{2} \right) dydz = 0,\label{ser8}
\end{eqnarray}
and it leads to the following quantities
\begin{gather}
X=-\chi \left( \beta(t) {\psi} +\mu(t)\overline{\sigma} \right), \label{ser9}\\
Y=-D\left (\frac{\beta(t) \overline{\sigma}}{2y} + \mu(t) \overline{\xi}+\alpha(t) \right ). \label{ser10}
\end{gather}
Inserting expressions (\ref{ser9}) and (\ref{ser10}) into Eqs.~(\ref{ser1}) and (\ref{ser2}), we obtain the relaxation equations
\begin{gather}
\frac{\partial \overline{\xi}}{\partial t}+{\bf u}\cdot \nabla\overline{\xi}=\frac{\partial}{\partial z} \left( \frac{\overline{\sigma}^2}{4y^2} \right) -\chi (\beta(t) {\psi}+ \mu(t) \overline{\sigma}),
 \label{ser15} \\
\frac{\partial \overline{\sigma}}{\partial t}+{\bf u}\cdot \nabla\overline{\sigma}=-D \left ( \frac{\beta(t)\overline{\sigma}}{2y} +\mu(t) \overline{\xi}+\alpha(t) \: \right ).\label{ser16}
\end{gather}
The Lagrange multipliers evolve so as to satisfy the constraints. Substituting Eqs. (\ref{ser9}) and (\ref{ser10}) in Eqs. (\ref{ser5}) and (\ref{ser6}), and recalling Eq. (\ref{ser4}), we obtain the algebraic equations
\begin{equation}
\frac{1}{\beta(t)}=2E^{f.g.}-\langle\overline{\xi}\psi\rangle-\left\langle \frac{\overline{\sigma}^2}{2y}\right\rangle,\label{ser11}
\end{equation}
\begin{eqnarray}
\left (\left\langle\chi\psi\overline{\sigma}\right\rangle + \left\langle D\frac{\overline{\sigma}\overline{\xi}}{2y}\right\rangle\right )\beta(t)+\left (\left\langle\chi\overline{\sigma}^2\right\rangle
+\left\langle D{\overline{\xi}^2}\right\rangle\right )\mu(t)\nonumber\\
 + \left\langle D\overline{\xi}\right\rangle\alpha(t)
 =0,\qquad\label{ser12}
\end{eqnarray}
\begin{eqnarray}
\left\langle D\frac{\overline{\sigma}}{2y}\right\rangle\beta(t)+\left \langle D\overline{\xi}\right\rangle\mu(t)+\alpha(t)\langle D\rangle
 =0.\qquad\label{ser13}
\end{eqnarray}
Substituting $\psi$ and $\overline{\sigma}/y$ taken from Eqs. (\ref{ser9}) and (\ref{ser10}) in Eq. (\ref{ser3}) and using the constraints (\ref{ser5}) and (\ref{ser6}), we easily obtain
 \begin{eqnarray}
\dot S=\int \frac{X^2}{\chi}\, dydz+\int \frac{Y^2}{D}\, dydz,\label{ser14}
\end{eqnarray}
so that $\dot S\ge 0$ provided that $D$ and $\chi$ are both positive. On the other hand $\dot S=0$ iff $X=Y=0$ leading to the conditions of equilibrium  given by Eqs. (\ref{ses15}) and (\ref{ses16}). From Lyapunov's direct method, we conclude that these relaxation equations tend to a maximum of entropy at fixed microscopic energy, helicity and angular momentum.

The relaxation equations (\ref{ser15}) and (\ref{ser16}) are similar to Eqs. (\ref{ber19}) and (\ref{ber20}) but the constraints determining the evolution of the Lagrange multipliers are different. More precisely,  Eqs. (\ref{ser12}) and (\ref{ser13}) coincide with Eqs. (\ref{ber15}) and (\ref{ber16}) but  Eq. (\ref{ber14}) has been replaced by Eq. (\ref{ser11}).   Indeed, in the present approach, the distribution of angular momentum is always Gaussian during the dynamical evolution. It is given by Eq. (\ref{ses2}) at any time, {\it i.e.}
\begin{equation}
\rho({\bf r},\eta,t) = \left (\frac{\beta(t)}{4\pi y}\right )^{1/2} e^{-\frac{\beta(t)}{4y}(\eta-\overline{\sigma}({\bf r},t))^2}.
\end{equation}
By contrast, in Sec. \ref{BasicPrin}, the distribution of angular momentum changes with time. Therefore, the dynamical evolution is different. However, in the two approaches, the equilibrium state is the same, {\it i.e.} it solves the maximization problem  (\ref{bes1}). This is sufficient if we use these  relaxation equations as numerical algorithms to compute the maximum entropy state.

{\it Remark:} Using Eqs. (\ref{ser1}), (\ref{ser2}) and (\ref{ser3}), it is easy to show that $\dot S=-\beta(t)\dot{E}^{c.g.}$ so that $\dot{E}^{c.g.}\le 0$ since $\beta(t)\ge 0$. Therefore, the macroscopic energy monotonically decreases through the relaxation equations. This is to be expected since the maximization problem (\ref{ses9})  is equivalent to the minimization of the macroscopic energy at fixed helicity and angular momentum (see Sec. \ref{sec_demin}).

\subsection{Relaxation equations associated with the minimization problem (\ref{ses9b})}
\label{sec_relaxbel}

We shall introduce a set of relaxation equations associated with the minimization problem (\ref{ses9b}). We write the dynamical equations as
\begin{gather}
\frac{\partial \overline{\xi}}{\partial t}+{\bf u}\cdot \nabla\overline{\xi}=\frac{\partial}{\partial z} \left( \frac{\overline{\sigma}^2}{4y^2} \right)+X, \label{rer1} \\
\frac{\partial\overline{\sigma}}{\partial t}+{\bf u}\cdot \nabla\overline{\sigma}=Y,\label{rer2}
\end{gather}
where $X$ and $Y$ are two unknown quantities to be chosen so as to decrease $E^{c.g.}$ while conserving  $H$ and $I$. The time variations of $E^{c.g.}$ are given by
\begin{equation}
\dot{E}^{c.g.}=\int \psi X\, dydz+\int \frac{\overline{\sigma}}{2y}Y\, dydz.\label{rer2b}
\end{equation}
On the other hand, the time variations of $H$ and $I$ are
\begin{gather}
\dot{{H}}=0=\int X \overline{\sigma} \, dydz+ \int Y\overline{\xi} \, dydz,\label{rer3}\\
\dot{{I}}=0=\int Y dydz.\label{rer4}
\end{gather}

Following the Maximum Entropy Production Principle, we maximize the dissipation $\dot{E}^{c.g.}$ with $\dot{{I}}=\dot{{H}}=0$ and the additional constraints
\begin{equation}
\frac{X^2}{2}\le C_\xi({\bf r},t), \quad \frac{Y^2}{2}\le C_{\sigma}({\bf r},t).\label{rer4b}
\end{equation}
The variational principle can be written in the form
\begin{eqnarray}
\delta \dot{E}^{c.g.} + \mu(t) \delta \dot{{H}} + \alpha(t) \delta \dot{{I}} \nonumber \\
+\int \frac{1}{\chi({\bf r},t)} \delta \left( \frac{X^2}{2} \right) dydz\nonumber \\
+\int \frac{1}{D({\bf r},t)} \delta \left(\frac{Y^2}{2} \right) dydz  = 0,\label{rer5}
\end{eqnarray}
and we obtain the  following quantities
\begin{gather}
X=-\chi \left({\psi} +\mu(t)\overline{\sigma} \right), \label{rer6}\\
Y=-D\left (\frac{\overline{\sigma}}{2y} + \mu(t) \overline{\xi}+\alpha(t) \right ). \label{rer7}
\end{gather}
Substituting Eqs. (\ref{rer6}) and (\ref{rer7}) into Eq.~(\ref{rer1}) and (\ref{rer2}) leads to the following relaxation equations
\begin{gather}
\frac{\partial \overline{\xi}}{\partial t}+{\bf u}\cdot \nabla\overline{\xi}=\frac{\partial}{\partial z} \left( \frac{\overline{\sigma}^2}{4y^2} \right) -\chi ({\psi}+ \mu(t) \overline{\sigma}),
 \label{rer11} \\
\frac{\partial \overline{\sigma}}{\partial t}+{\bf u}\cdot \nabla\overline{\sigma}=-D \left ( \frac{\overline{\sigma}}{2y} +\mu(t) \overline{\xi}+\alpha(t) \: \right ).\label{rer12}
\end{gather}
The Lagrange multipliers $\mu(t)$ and $\alpha(t)$ evolve so as to satisfy the constraints (\ref{rer3}) and  (\ref{rer4}). Substituting   Eqs. (\ref{rer6}) and  (\ref{rer7}) in Eqs. (\ref{rer3}) and (\ref{rer4}), we obtain the algebraic equations
\begin{eqnarray}
\left (\left\langle\chi\overline{\sigma}^2\right\rangle
+\left\langle D{\overline{\xi}^2}\right\rangle\right )\mu(t)
+ \left\langle D\overline{\xi}\right\rangle\alpha(t)\nonumber\\
+\left\langle\chi\psi\overline{\sigma}\right\rangle + \left\langle D\frac{\overline{\sigma}\overline{\xi}}{2y}\right\rangle=0,\label{rer8}
\end{eqnarray}
\begin{eqnarray}
\left\langle D\overline{\xi}\right\rangle\mu(t)+\alpha(t)\langle D\rangle+\left\langle D\frac{\overline{\sigma}}{2y}\right\rangle=0.
\label{rer9}
\end{eqnarray}
Substituting $\psi$ and $\overline{\sigma}/y$ taken from Eqs. (\ref{rer6}) and (\ref{rer7}) in Eq. (\ref{rer2b}) and using the constraints (\ref{rer3}) and (\ref{rer4}), we easily obtain
 \begin{eqnarray}
\dot{E}^{c.g.}=-\int \frac{X^2}{\chi}\, dydz-\int \frac{Y^2}{D}\, dydz,\label{rer10}
\end{eqnarray}
so that $\dot{E}^{c.g.}\le 0$ provided that $D$ and $\chi$ are both positive. On the other hand, $\dot{E}^{c.g.}=0$ iff $X=Y=0$ leading to the conditions of equilibrium  given by Eqs. (\ref{res16}) and (\ref{res17}). By Lyapunov's direct method, we conclude that these relaxation equations tend to a minimum of macroscopic energy $E^{c.g.}$ at fixed helicity and angular momentum. Therefore, the relaxation equations (\ref{rer11},\ref{rer12})  can be used as a numerical algorithm to solve the minimization problem (\ref{ses9b}).

{\it Remark:} since these relaxation equations solve Eq. (\ref{bel1}), they can also be used as a numerical algorithm to construct nonlinearly dynamically stable stationary solutions of the axisymmetric Euler equations corresponding to Beltrami states (see Secs. \ref{Eulersec} and \ref{bel}) independently of the statistical mechanics interpretation.

\section{Another type of relaxation equations}
\label{sec_other}

In the main part of the paper, we have not taken into account the conservation of circulation $\Gamma=\int \xi\, dydz$ because there is no critical point of energy at fixed helicity, angular momentum and circulation (see \cite{naso2}). Nevertheless, at the level of the relaxation equations, it is possible to take this constraint into account. We shall introduce a set of relaxation equations that minimize the energy $E^{c.g.}$ at fixed helicity $H$, angular momentum $I$ and circulation $\Gamma$. Since there is no energy minimum (not even a critical point of energy), these equations should have a non-trivial behavior. To derive these equations, one possibility is to write them in the form  (\ref{rer1})-(\ref{rer2}) and introduce Lagrange multipliers for each constraint. Another possibility is to write them in the form
\begin{gather}
\frac{\partial \overline{\xi}}{\partial t}+{\bf u}\cdot \nabla\overline{\xi}=\frac{\partial}{\partial z} \left( \frac{\overline{\sigma}^2}{4y^2} \right)-\nabla\cdot {\bf J}_\xi, \label{rer1x} \\
\frac{\partial\overline{\sigma}}{\partial t}+{\bf u}\cdot \nabla\overline{\sigma}=-\nabla\cdot {\bf J}_\sigma,\label{rer2x}
\end{gather}
where ${\bf J}_\xi$ and ${\bf J}_\sigma$ are two unknown currents to be chosen so as to decrease $E^{c.g.}$ while conserving  $H$. The form (\ref{rer1x})-(\ref{rer2x}) guarantees the conservation of circulation and angular momentum. The time variations of $E^{c.g.}$ are given by
\begin{equation}
\dot{E}^{c.g.}=\int {\bf J}_\xi\cdot \nabla\psi \, dydz+\int {\bf J}_\sigma\cdot \nabla\left (\frac{\overline{\sigma}}{2y}\right )\, dydz.\label{rer2bx}
\end{equation}
On the other hand, the time variations of $H$ are
\begin{gather}
\dot{{H}}=0=\int {\bf J}_\xi\cdot \nabla\overline{\sigma} \, dydz+ \int {\bf J}_\sigma\cdot\nabla \overline{\xi} \, dydz.\label{rer3x}
\end{gather}

Following the Maximum Entropy Production Principle, we maximize the dissipation $\dot{E}^{c.g.}$ with $\dot{{H}}=0$ and the additional constraints
\begin{equation}
\frac{{\bf J}_\xi^2}{2}\le C_\xi({\bf r},t), \quad \frac{{\bf J}_\sigma^2}{2}\le C_{\sigma}({\bf r},t).\label{rer4bx}
\end{equation}
The variational principle can be written in the form
\begin{eqnarray}
\delta \dot{E}^{c.g.} + \mu(t) \delta \dot{{H}}
+\int \frac{1}{D_\xi({\bf r},t)} \delta \left( \frac{{\bf J}_\xi^2}{2} \right) dydz\nonumber \\
+\int \frac{1}{D_\sigma({\bf r},t)} \delta \left(\frac{{\bf J}_\sigma^2}{2} \right) dydz  = 0,\label{rer5x}
\end{eqnarray}
and we obtain the optimal currents
\begin{gather}
{\bf J}_\xi=-D_\xi \left(\nabla{\psi} +\mu(t)\nabla\overline{\sigma} \right), \label{rer6x}\\
{\bf J}_\sigma=-D_\sigma\left \lbrack \nabla\left (\frac{\overline{\sigma}}{2y}\right ) + \mu(t) \nabla\overline{\xi}\right\rbrack. \label{rer7x}
\end{gather}
Substituting Eqs. (\ref{rer6x}) and (\ref{rer7x}) into Eq.~(\ref{rer1x}) and (\ref{rer2x}) leads to the following relaxation equations
\begin{gather}
\frac{\partial \overline{\xi}}{\partial t}+{\bf u}\cdot \nabla\overline{\xi}=\frac{\partial}{\partial z} \left( \frac{\overline{\sigma}^2}{4y^2} \right) +\nabla\cdot \left\lbrack D_\xi \left(\nabla{\psi} +\mu(t)\nabla\overline{\sigma} \right)\right\rbrack,
 \label{rer11x} \\
\frac{\partial \overline{\sigma}}{\partial t}+{\bf u}\cdot \nabla\overline{\sigma}=\nabla\cdot\left\lbrace  D_\sigma\left \lbrack \nabla\left (\frac{\overline{\sigma}}{2y}\right ) + \mu(t) \nabla\overline{\xi}\right\rbrack\right\rbrace.\label{rer12x}
\end{gather}
The Lagrange multiplier $\mu(t)$  evolves so as to satisfy the constraint (\ref{rer3x}). Substituting   Eqs. (\ref{rer6x}) and  (\ref{rer7x}) in Eq. (\ref{rer3x}), we obtain
\begin{eqnarray}
\mu(t)=-\frac{\int D_\xi\nabla\psi\cdot\nabla\overline{\sigma}\, dydz+\int D_\sigma \nabla\left (\frac{\overline{\sigma}}{2y}\right )\cdot\nabla\overline{\xi}\, dydz}{\int D_\xi (\nabla\overline{\sigma})^2\, dydz+\int D_\sigma (\nabla\overline{\xi})^2\, dydz}.\nonumber\\
\label{rer8x}
\end{eqnarray}
Substituting $\nabla\psi$ and $\nabla(\overline{\sigma}/y)$ taken from Eqs. (\ref{rer6x}) and (\ref{rer7x}) in Eq. (\ref{rer2bx}) and using the constraint (\ref{rer3x}), we easily obtain
 \begin{eqnarray}
\dot{E}^{c.g.}=-\int \frac{{\bf J}_\xi^2}{D_\xi}\, dydz-\int \frac{{\bf J}_\sigma^2}{D_\sigma}\, dydz,\label{rer10x}
\end{eqnarray}
so that $\dot{E}^{c.g.}\le 0$ provided that $D_\xi$ and $D_\sigma$ are both positive.

%
%

\end{document}